\documentclass[12pt,psf,epsf]{article}
\usepackage[centertags]{amsmath}
\usepackage{amssymb}
\usepackage{graphicx}
\usepackage{epsfig}
\usepackage{ulem}
\usepackage[english]{babel}
\usepackage{array}
\usepackage{amsthm}
\usepackage{latexsym}
\usepackage[mathcal]{euscript}
\pdfoutput=1
\usepackage{epsfig}
 \usepackage{jheppubm}
\usepackage{mathdots}
\usepackage{MnSymbol}
\usepackage{multirow}

\newcommand{\blue}{\textcolor{blue}}
\newcommand{\nn}{\nonumber}
\newcommand{\be}{\begin{equation}}
\newcommand{\ee}{\end{equation}}
\newcommand{\bea}{\begin{eqnarray}}
\newcommand{\eea}{\end{eqnarray}}

\newcommand{\e}{\mathrm{e}}
\newcommand{\ZZ}{\mathbb{Z}}

\newcommand{\dd}{\partial}

\newcommand{\bj}{{\bf j}}

\DeclareMathOperator{\Pf}{Pf}

\DeclareMathOperator{\sgn}{sgn}

\newcommand{\mZ}{\mathcal{Z}}
\newcommand{\mG}{\mathcal{G}}

\newcommand{\fl}{\mathfrak{l}}

\newcommand{\fm}{\mathfrak{m}}

\definecolor{dgreen}{rgb}{0.,0.6,0.}

\title{Revealing nonperturbative effects in the SYK model}
\author{Irina Aref'eva, Mikhail Khramtsov, Igor Volovich}

\affiliation{Steklov Mathematical Institute, Russian Academy of Sciences,\\Gubkina str. 8, 119991, Moscow, Russia}

\abstract{We study the large $N$ saddle points of two SYK chains coupled by an interaction that is nonlocal in Euclidean time. We start from analytic treatment of the free case with $q=2$ and perform the numerical study of the interacting case $q=4$. We show that in both cases there is a nontrivial phase structure with infinite number of phases. Every phase correspond to a saddle point in the non-interacting two-replica SYK. The nontrivial saddle points have non-zero value of the replica-nondiagonal correlator in the sense of quasi-averaging, when the coupling between replicas is turned off. Thus, the nonlocal interaction between replicas provides a protocol for turning the nonperturbatively subleading effects in SYK into non-equilibrium configurations which dominate at large $N$. For comparison we also study two SYK chains with local interaction for $q=2$ and $q=4$. We show that the $q=2$ model also has a similar phase structure, whereas in the $q=4$ model, dual to the traversable wormhole, the phase structure is different.}

\emailAdd{arefeva@mi-ras.ru}
\emailAdd{khramtsov@mi-ras.ru}
\emailAdd{volovich@mi-ras.ru}

\begin{document}
\maketitle

\section{Introduction}
\label{sec:Intro}

The SYK model \cite{Sachdev92,Kitaev,MScomments,Kitaev17} is an example of maximally chaotic holographic quantum theory in $0+1$ dimensions that is solvable at large $N$. An important feature of the SYK model is that the exact path integral can be formulated in terms of bilocal collective fields \cite{Sachdev15,Kitaev,Jevicki16}. This framework allows for performing the $1/N$-expansion systematically as the saddle point expansion.

The bulk theory dual to SYK contains the $2$D Jackiw-Teitelboim theory \cite{Jackiw,Teitelboim,Almheiri14,Engelsoy16,Maldacena16,Jensen16} as the gravity subsector. More specifically, the holographic correspondence dictates that the partition function of $M$ copies (replicas) of SYK should be equal to the partition function of the bulk theory, which includes the JT gravity on constant negative curvature spacetimes with $M$ boundaries \cite{Cotler16,Saad18,Saad19,Blommaert18,Blommaert19,Maldacena18}. Therefore, one can think of collective replica field formulation of SYK as a representation for the bulk theory path integral. From this point of view, the saddle points of the path integral correspond to some bulk spacetime geometries of constant negative curvature with $M$ boundaries, and the perturbative contributions to the $1/N$ expansion come from quantum gravity effects on a fixed geometry saddle point. 

The recent studies of quantum chaotic nature of SYK-type theories and random matrices \cite{Cotler16,Garcia-garcia16,Saad18,Garcia-Garcia19,Gur-Ari18,Lau18,Cotler17, Gharibyan18,Okuyama18,Okuyama19,Garcia-Garcia17,Balasubramanian16} demonstrate that the nonperturbative effects in the $1/N$ expansion of SYK-type models play essential role in manifestations of quantum chaos and random matrix universality. In particular, it was shown that the replica-nondiagonal saddle points of the replica field path integral are related to the ramp behavior of the spectral form-factor \cite{Saad18}. Replica-nondiagonal structures of saddle points of SYK-like models were also studied in \cite{Georges00,Arefeva18,Arefeva19,Fu16,Gur-Ari18,AKTV,Kamenev}, in particular in relation to the issue of spin glasses and replica symmetry breaking. From the gravity point of view, the nonperturbative effects arising from the subleading saddle points are related to the topologies of higher genus \cite{Saad19,Harlow18,Blommaert19}, and they probe discrete spectrum of black hole microstates and are important for the recovery of lost information \cite{Cotler16,Saad18,Saad19}.

In the previous work \cite{AKTV} we have obtained a family of exact replica-nondiagonal saddle points in the SYK model with $M$ replicas, where $M$ is a positive integer number. We have shown that these saddle points give nonperturbative contributions to the path integrals, that are suppressed at large $N$ as $\e^{-N \delta I}$ compared to the replica-diagonal saddle point. In the current paper we study the model of $M=2$ coupled SYK replicas with interaction, which turns those saddle points into dominant ones. We treat the quadratic SYK$_2$ case analytically and the case of quartic interaction SYK$_4$ numerically. In both cases the interaction between replicas generates a nontrivial phase structure with infinite number of phases. The order parameter for these phases is the replica-offdiagonal correlator. We show that on non-trivial phases it acquires non-zero value in the sense of Bogolyubov quasi-averaging, matching the corresponding subleading saddle point in the SYK with the interaction turned off. 

The paper is organized as follows. In the section \ref{sec:Setup} we introduce the model of two SYK chains with nonlocal coupling. The section \ref{sec:q=2} is devoted to the study of large $N$ saddle points and phase structure in the integrable $q=2$ variant of the model. In the section \ref{sec:q=4} we perform the numerical study of the phase structure in the $q=4$ version of the model, using the results of the previous section for initial approximation to numerical solution of saddle point equations. In the section \ref{sec:Wormholes} we discuss the model with local coupling that is related to the two-dimensional traversable wormhole \cite{Maldacena17,Maldacena18,Garcia-Garcia19}. We also study saddle points in $q=2$ and $q=4$ versions of the model and compare it to our model. In the section \ref{sec:Symmetry} we establish the connection between the interaction between SYK replicas and spontaneous symmetry breaking in the sense of Bogolyubov quasi-averages in replica-nondiagonal saddle points of SYK. Finally, the section \ref{sec:Discussion} is devoted to discussion of the results and open questions.

\section{The model of two nonlocally interacting SYK chains}
\label{sec:Setup}

We study the model of $2 N$ Majorana fermions $\psi_i^\alpha$ in $0+1$ dimensions. Here the replica index $\alpha = L,R$, and $i = 1, \dots, N$ is the color (or site) index. The two replicas have disordered $q$-fermion interactions with Gaussian randomized couplings ${\bf j} = \{ j_{i_1 \dots i_q} \}$, equal between replicas for every realization of the randomness. 

The total action for the fermions is given by\footnote{In this paper we work in Euclidean signature exclusively.} 
\be
S[\psi, {\bf j}] = S_L[\psi^L, {\bf j}] + S_R[\psi^R, {\bf j}] + S_{\text{int}}[\psi]\,. \label{Stotal}
\ee
Here 
\be
S_\alpha[\psi^\alpha, {\bf j}] = \int_0^\beta d\tau  \left(- \frac12 \sum_{i=1}^N \psi_i^\alpha \frac{d}{d \tau} \psi_i^\alpha -  \frac{i^{q/2}}{q!} \sum_{i_1, i_2,\dots,i_q=1}^N j_{i_1 i_2\dots\i_q} \psi_{i_1}^\alpha \psi_{i_2}^\alpha \dots \psi_{i_q}^\alpha\right) \label{Ssyk} 
\ee
is the single-replica SYK$_q$ action. The number $q$ is the order of the fermionic interaction. In the present work we consider cases $q=2$ and $q=4$.

We also have the term with the interaction between replicas, which has the nonlocal form:
\be
S_{\text{int}} = \int_0^\beta \int_0^\beta d\tau_1 d\tau_2 \sum_{i=1}^N \psi_i^L (\tau_1) \eta_{LR} (\tau_1 - \tau_2) \psi_i^R(\tau_2)\,. \label{Sint}
\ee
We choose the interaction $\eta$ such that at finite temperature $\beta^{-1}$ it has the form 
\be
\eta_{LR} (\tau_1 - \tau_2) = \eta_{RL} (\tau_1 - \tau_2) = \frac{\nu}{\beta \sin \frac{\pi}{\beta} (\tau_1 - \tau_2)}\,. \label{eta_sym(t)T}
\ee
At zero temperature we have 
\be
\eta_{LR} (\tau_1 - \tau_2) = \eta_{RL} (\tau_1 - \tau_2) = \frac{\nu}{\pi (\tau_1 - \tau_2)}\,. \label{eta_sym(t)}
\ee
We choose the interaction of this form, because in the frequency space in both cases (\ref{eta_sym(t)}),(\ref{eta_sym(t)T}) has a simple form: 
\be
\eta_{LR} (\omega) = \eta_{RL} (\omega) =  i \nu \sgn(\omega)\,. \label{zeta_sym}
\ee
Indeed, 
\be
\frac{i}{2\pi} \int_{-\infty}^{+\infty}  \sgn(\omega) \e^{- i \omega \tau} d\omega = \frac{1}{\pi \tau}\,, 
\ee
where the integral is understood in the sense of principal value. At finite temperature frequency is quantized as follows:
\be
\omega_n = \frac{2\pi}{\beta} \left(n + \frac12\right)\,, \quad n \in \ZZ\,, \label{Matsubara}
\ee
so in this case the Fourier transform reads 
\be
\frac{i}{\beta} \sum_{n = -\infty}^{+\infty} \sgn(\omega_n) \e^{-i \omega_n \tau} = \frac{1}{\beta}\frac{1}{\sin \frac{\pi \tau}{\beta}}\,.
\ee
Note that we also have to understand the integral (\ref{Sint}) in the regularized sense (\blue{principal value}). The constant $\nu$ has the dimension of energy and determines the strength of the interaction between replicas. We note that for $q=2$ the action (\ref{Sint}) looks similar to the single-replica nonlocal kinetic term, considered in \cite{Gross17}. On the other hand, in the previous work \cite{Maldacena18,Saad18,Kim19,Garcia-Garcia19,Maldacena17}, where coupled replicas were considered, the interaction was local.

The partition function is given by 
\be
Z_{\bj}(\beta) = \int D\psi \e^{-S[\psi, \bj]}\,.
\ee
By performing the standard procedure of averaging over the Gaussian disorder and integrating out the fermions (see \cite{Kitaev17} for detailed derivation), one obtains the path integral over the bilocal collective replica fields $G_{\alpha\beta}(\tau_1, \tau_2)$, $\Sigma_{\alpha\beta}(\tau_1, \tau_2)$
\bea
\mZ(\eta) &=& \int DG D\Sigma\ \e^{-N I[G, \Sigma]}\,; \label{Z(eta)}\\
 I[G, \Sigma; \eta] &=& - \log \Pf[\delta_{\alpha\beta} \dd_\tau - \hat{\Sigma}_{\alpha\beta}] +\frac12 \int_0^\beta \int_0^\beta d\tau_1 d \tau_2  \left(\Sigma_{\alpha\beta}(\tau_1, \tau_2) G_{\alpha\beta}(\tau_1, \tau_2) - \frac{J^2}{q} G_{\alpha\beta}(\tau_1, \tau_2)^q\right) \nn\\&&-\frac{1}{2} \int_0^\beta \int_0^\beta d \tau_1 d\tau_2 \eta_{\alpha\beta}(\tau_1, \tau_2) G_{\alpha\beta}(\tau_1, \tau_2)\,, \label{I(eta)}
\eea
The bilocal fields are supposed to satisfy the antisymmetry condition
\be
G_{\alpha\beta}(\tau_1, \tau_2) = - G_{\beta\alpha}(\tau_2, \tau_1)\,; \qquad \Sigma_{\alpha\beta}(\tau_1, \tau_2) = - \Sigma_{\beta\alpha}(\tau_2, \tau_1)\,.\label{antisymmetry}
\ee
The interaction between replicas of the form (\ref{Sint}), which is bilinear in fermions, is equivalent to inclusion of the source term for the field $G$\footnote{An equivalent (up to a straightforward redefinition of field variables) way to think about $\eta$ is as shift of the $\Sigma$ field \cite{Arefeva18,Arefeva19}.}. More specifically, we have 
\be
\eta(\tau_1, \tau_2) = \left(\begin{matrix}
	0 & \zeta(\tau_1-\tau_2) & \\
	\zeta(\tau_1-\tau_2) & 0 &
\end{matrix}\right)\,, \label{eta_sym}
\ee
where the function $\zeta$ has the form (\ref{zeta_sym}) in the frequency space and (\ref{eta_sym(t)}) or (\ref{eta_sym(t)T}) in the coordinate space. Note that this is a symmetric matrix in the replica space. 

Assuming $\nu$ is constant, the Fourier transform to the temporal representation in the case of zero temperature is given by the formula (\ref{eta_sym(t)}), and at finite temperature by (\ref{eta_sym(t)T}). Both those expressions define integral kernels of antisymmetric operators in the space of functions of time. The combined symmetry properties of the source are in agreement with the general antisymmetry condition
\be
\eta_{LR}(\tau_1, \tau_2) = - \eta_{RL}(\tau_2, \tau_1)\,. \label{eta_antisymmetry}
\ee

We are interested in the study of saddle points of the path integral (\ref{Z(eta)}). The saddle point equations read
\bea
&&  \dd_\tau G_{\alpha\gamma}(\tau, \tau'')-\int d\tau' G_{\alpha\beta}(\tau, \tau') \Sigma_{\beta\gamma}(\tau', \tau'') =  \delta_{\alpha\gamma}\delta(\tau-\tau'')\,; \label{saddle-point-1-eta}\\
&& \Sigma_{\alpha\beta}(\tau, \tau') = J^2 G_{\alpha\beta}(\tau, \tau')^{q-1} + \eta_{\alpha\beta}(\tau, \tau') \,. \label{saddle-point-2-eta}
\eea
The replica-symmetric form of the source dictates via the equation (\ref{saddle-point-2-eta}), that for solutions we would want to assume the replica-symmetric form as well: 
\be\label{NDS}
G_{LL}=G_{RR}=G_{0},\,\,\,\,\,G_{LR} = G_{RL}=G_{1}\,,
\ee
where $G_0$ and $G_1$ are both assumed to be odd functions in the frequency and time representations, and to be antiperiodic in time. In \cite{AKTV} the constructed solutions for non-interacting replica case had the same assumptions applied. 

To conclude this section, let us make some remarks for future reference: 

\begin{itemize}
	\item Another possibility to satisfy the condition (\ref{eta_antisymmetry}) is to consider the case when $\eta$ is antisymmetric in the replica space and symmetric in the coordinate space. This case is widely considered in literature, in particular in the context of traversable wormholes and chaotic thermofield double dynamics \cite{Saad18,Maldacena17,Maldacena18,Blommaert19} and $O(N)$ symmetry breaking \cite{Kim19}. We will consider it separately in the section \ref{sec:Wormholes}.
	
	\item An important assumption about the source $\eta$, that we make, is that it depends only on difference of times. Thus we only focus on solutions that are translationally invariant with respect to simultaneous translation in both replicas, breaking a part of the total translational symmetry. We discuss this in section \ref{sec:Symmetry}. 
\end{itemize}

\section{Quadratic case}
\label{sec:q=2}

\subsection{Equations of motion}

We start the study of saddle points and phase structure with the case $q=2$, which corresponds to the integrable variant of the SYK model, or random mass fermions. We look for the solutions of equations (\ref{saddle-point-1-eta})-(\ref{saddle-point-2-eta}). Assuming the ansatz for $G$ of the form (\ref{NDS}), the saddle point equation (\ref{saddle-point-2-eta}) has the form in the frequency space
\bea
\Sigma_0(\omega) &=& J^2 G_0(\omega)\,; \\
\Sigma_1(\omega) &=& J^2 G_1(\omega) + i \nu \sgn (\omega) \,.
\eea
Substituting into the equation (\ref{saddle-point-1-eta}), we get a system of two algebraic equations for two unknowns:
\bea\label{eq1-nu}
-i\omega G_0(\omega) -J^2 G_0(\omega)^2- (J^2 G_1(\omega)+i \nu \sgn (\omega)) G_1(\omega) &=&1\,; \\
\label{eq2-nu}-i \omega G_1(\omega) -J^2 G_0(\omega) G_1(\omega) - (J^2 G_1(\omega) +i \nu \sgn (\omega)) G_0(\omega) &=&0 \,.
   \eea
This form of equations is valid for both zero and finite temperature cases for $q=2$. In the case $\nu = 0$ one recovers the equations for the decoupled replicas for $M=2$, which were solved in \cite{AKTV}. 

Before we proceed to discussion of the solutions, let us note that the peculiar feature of the $q=2$ case is that the source $\eta$ in the form (\ref{zeta_sym}) respects the conformal symmetry in the IR limit and transforms with fixed conformal dimension $\Delta=\frac12$. The same form of the source breaks conformal symmetry in the IR limit in cases when $q > 2$.    

\subsection{Solutions}
\label{sec:q=2-solutions-sym}

Now let us discuss the solutions. We focus on the case of finite temperature. In this case the frequencies are quantized according to the Matsubara rule (\ref{Matsubara}). The equations (\ref{eq1-nu})-(\ref{eq2-nu}) for general value of $\nu$ admit $4$ solutions for every Matsubara frequency. They are given by the formulae
\bea
G_0^{(1)}(\omega_n) &=& -\frac{i}{2 J^2} \left[\omega_n - \frac{\sgn(\omega_n)}{\sqrt{2}}A_1\right]\,; \label{G01-nu}\\ 
G_1^{(1)}(\omega_n) &=& \frac{i}{8 J^2 \nu \omega_n} \left(\sqrt{2} A_1 (4 J^2 + \omega_n^2 - B) - \nu^2 (4 |\omega_n| - \sqrt{2} A_1) \right)\,;\label{G11-nu}\\
G_0^{(2)}(\omega_n) &=& -\frac{i}{2 J^2} \left[\omega_n + \frac{\sgn(\omega_n)}{\sqrt{2}} A_1  \right]\,;\label{G02-nu}\\
G_1^{(2)}(\omega_n) &=& \frac{i}{8 J^2 \nu \omega_n} \left(-\sqrt{2} A_1 (4 J^2 + \omega_n^2 - B) - \nu^2 (4 |\omega_n| + \sqrt{2} A_1) \right)\,;\label{G12-nu}
\eea
\bea
G_0^{(3)}(\omega_n) &=& -\frac{i}{2 J^2} \left[\omega_n + \frac{\sgn(\omega_n)}{\sqrt{2}}A_2\right]\,;\label{G03-nu}\\
G_1^{(3)}(\omega_n) &=& \frac{i}{8 J^2 \nu \omega_n} \left(-\sqrt{2} A_2 (4 J^2 + \omega_n^2 + B) - \nu^2 (4 |\omega_n| + \sqrt{2} A_2) \right)\,;\label{G13-nu}\\
G_0^{(4)}(\omega_n) &=& -\frac{i}{2 J^2} \left[\omega_n - \frac{\sgn(\omega_n)}{\sqrt{2}}A_2\right]\,;\label{G04-nu}\\
G_1^{(4)}(\omega_n) &=& \frac{i}{8 J^2 \nu \omega_n} \left(\sqrt{2} A_2 (4 J^2 + \omega_n^2 + B) - \nu^2 (4 |\omega_n| - \sqrt{2} A_2) \right)\,.\label{G14-nu}\\
\eea
Here we have introduced the auxiliary notations: 
\bea
A_1 &=& \sqrt{4 J^2 + \omega_n^2 + \nu^2 + B}\,;\\
A_2 &=& \sqrt{4 J^2 + \omega_n^2 + \nu^2 - B}\,;\\
B &=& \sqrt{16 J^2 \omega_n^2 + (4 J^2 + \nu^2 - \omega_n^2)^2}\,.
\eea
Let us expand these solutions in small $\nu$. Taking into account leading and subleading terms, we obtain 
\bea
G_0^{(1)}(\omega_n) &=& \frac{-i \omega_n + i \sgn(\omega_n) \sqrt{4J^2 + \omega_n^2}}{2J^2} + \frac{i \sgn(\omega_n)}{(4J^2 + \omega_n^2)^{\frac32}}\nu^2 + O(\nu^4)\,;\label{G01-nu-ser}\\ 
G_1^{(1)}(\omega_n) &=& \frac{i \omega_n - i \sgn(\omega_n) \sqrt{4J^2 + \omega_n^2}}{2J^2  \sqrt{4J^2 + \omega_n^2}} \nu - \frac{i \omega_n}{(4J^2 + \omega_n^2)^{\frac52}}\nu^2 + O(\nu^4)\,;\label{G11-nu-ser}\\
G_0^{(2)}(\omega_n) &=& \frac{-i \omega_n - i \sgn(\omega_n) \sqrt{4J^2 + \omega_n^2}}{2J^2} - \frac{i \sgn(\omega_n)}{(4J^2 + \omega_n^2)^{\frac32}}\nu^2 + O(\nu^4)\,;\label{G02-nu-ser}\\ 
G_1^{(2)}(\omega_n) &=& \frac{-i \omega_n - i \sgn(\omega_n) \sqrt{4J^2 + \omega_n^2}}{2J^2  \sqrt{4J^2 + \omega_n^2}} \nu + \frac{i \omega_n}{(4J^2 + \omega_n^2)^{\frac52}}\nu^2 + O(\nu^4)\,;\label{G12-nu-ser}\\
G_0^{(3)}(\omega_n) &=&  -\frac{i \omega_n}{2 J^2} - \frac{i \omega_n}{2J^2 \sqrt{4J^2 + \omega_n^2}}\nu + O(\nu^3)\,;\label{G03-nu-ser}\\
G_1^{(3)}(\omega_n) &=& -i \sgn(\omega_n) \frac{\sqrt{4J^2 + \omega_n^2}}{2J^2} - \frac{i \sgn(\omega_n)}{2J^2} \nu + O(\nu^3)\,.\label{G13-nu-ser}\\
G_0^{(4)}(\omega_n) &=& -\frac{i \omega_n}{2 J^2} + \frac{i \omega_n}{2J^2 \sqrt{4J^2 + \omega_n^2}}\nu + O(\nu^3)\,;\label{G04-nu-ser}\\
G_1^{(4)}(\omega_n) &=&i \sgn(\omega_n) \frac{\sqrt{4J^2 + \omega_n^2}}{2J^2} - \frac{i \sgn(\omega_n)}{2J^2} \nu + O(\nu^3)\,. \label{G14-nu-ser}\\
\eea
Comparing to the results of section 3.1 in \cite{AKTV} in the decoupled case $\nu=0$, we see that in the limit $\nu \to 0$ solutions 1 and 2 turn into the replica-diagonal solutions, whereas the solutions 3 and 4 turn into the replica-nondiagonal solutions. It is also important to note that for solutions 1 and 2 the subleading powers in $\nu$ do not change the leading UV asymptotic $\omega_n \to \infty$. 

The solutions of the full SYK$_2$ model are constructed by choosing each of the four roots for every Matsubara frequency independently. The solution, for which we choose the root 1 for every Matsubara frequency, gives the standard saddle point \cite{MScomments,Cotler16}. At $\nu \to 0$, it converges to the saddle point that dominates over saddle points, defined by other solutions \cite{AKTV}. In order to have the UV-finite action, we have to consider the solutions, for which only a finite number of Matsubara modes have roots 2, 3 or 4, and the UV asymptotic is defined by the solution 1 exclusively. 

\subsection{On-shell action and phase structure}
\label{sec:phases-q=2}

Now we discuss the contributions of saddle points to the path integral (\ref{Z(eta)}). To study the dominance of the saddle points, we consider the action density $\rho$, defined by the formula
\be
I[G, \Sigma]|_{\text{on-shell}} = \sum_n \rho(\omega_n, J, \nu)\,.\label{Ss}
\ee
The explicit expression for $\rho$ is obtained from (\ref{I(eta)}) by setting $q=2$, using the equation (\ref{saddle-point-2-eta}) and going into the frequency space. The action density is written as
\bea
\rho&=&-\fl+\frac{J^2}2\,\fm\label{sj}\,,
\eea 
where we have the contribution from the Pfaffian term 
\bea
\fl&=& \log\left[\left(1+\frac{J^2 G_0(\omega_n)}{i\omega_n}\right)^2 -\left(\frac{J^2 G_1(\omega_n) + i \nu \sgn(\omega_n)}{i \omega_n}\right)^2 \right]\,,
\eea
and the polynomial part 
\bea
\fm &=&  |G_0(\omega_n)|^2
+|G_1(\omega_n)|^2\,. \label{fm}
\eea
\begin{figure}[t]
	\centering
	\includegraphics[scale=0.28]{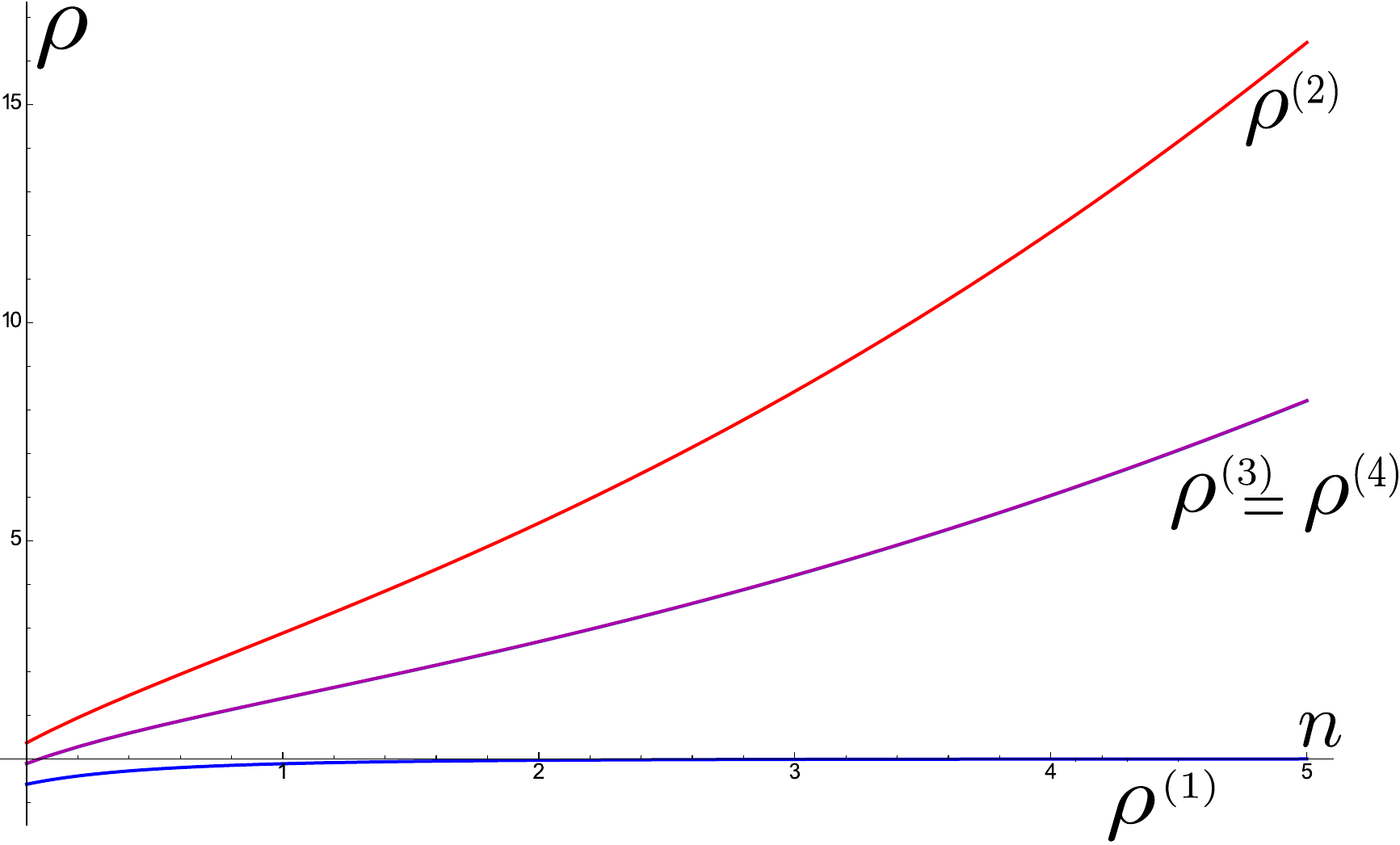}
	\includegraphics[scale=0.28]{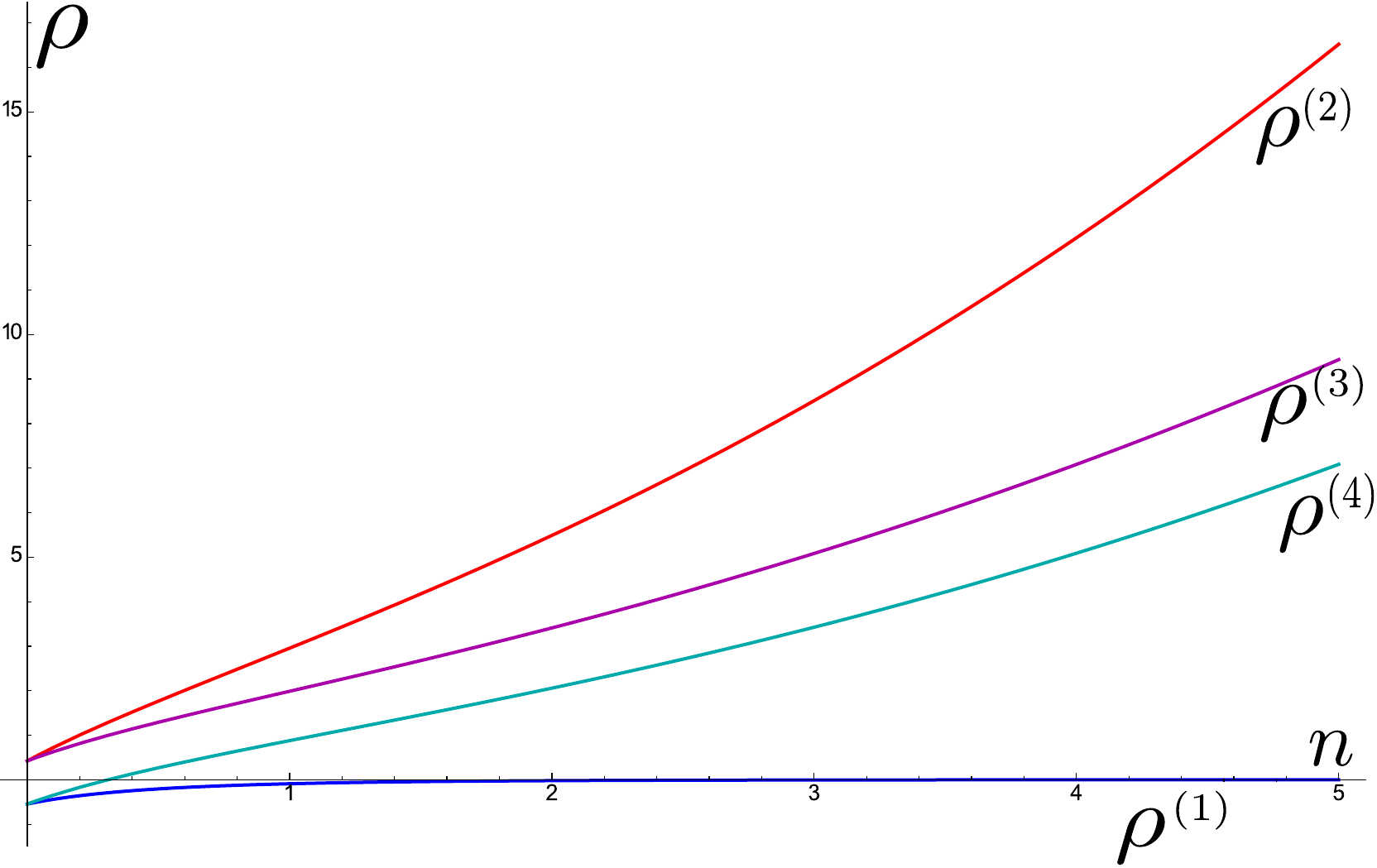}
	\includegraphics[scale=0.28]{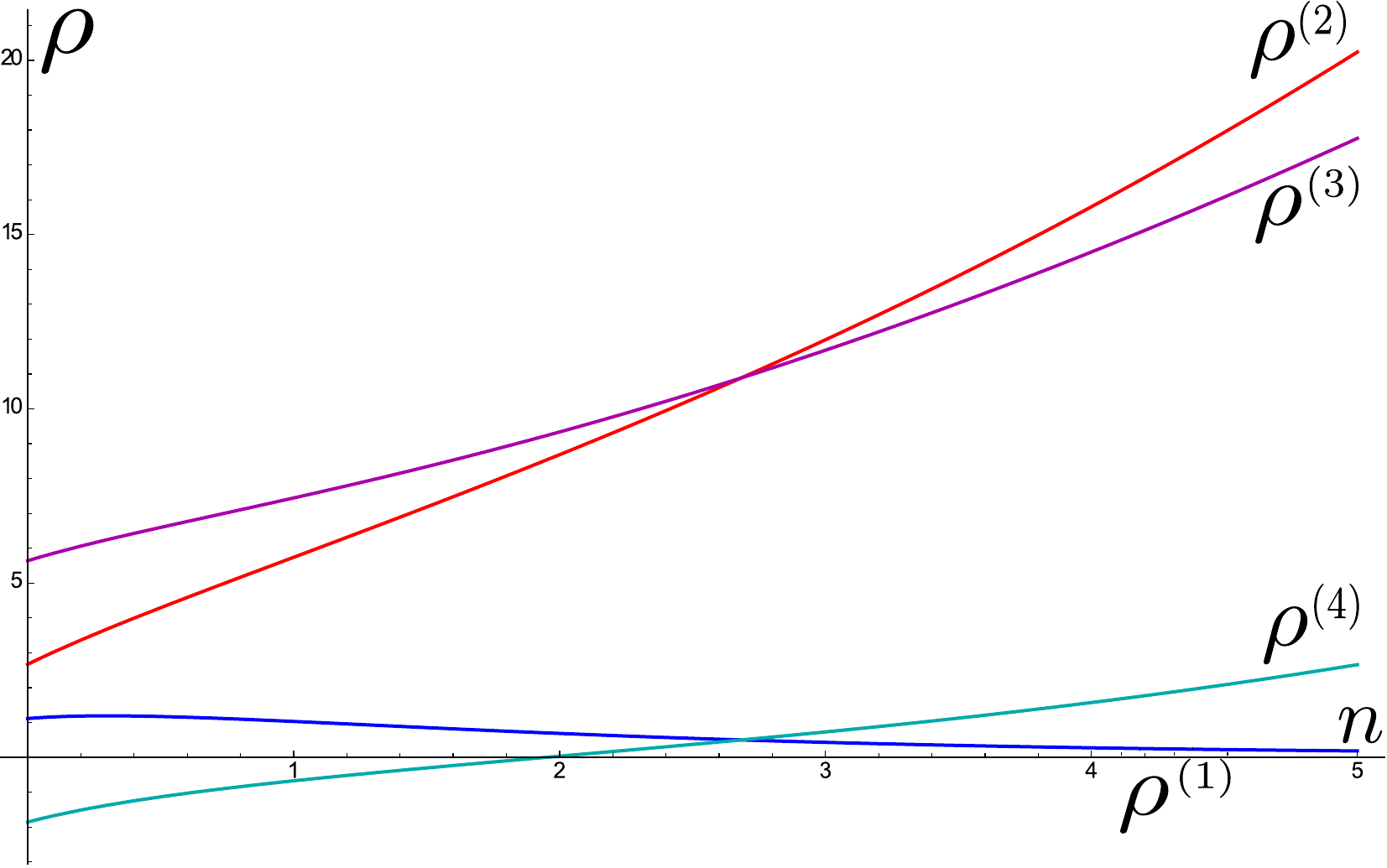} \\ 
	 (a) $\qquad \qquad \qquad \qquad\qquad $ (b) $ \qquad \qquad \qquad \qquad \qquad$ (c)
	\caption{Cartoon of action densities \eqref{action1-4-nu} as functions of Matsubara frequency label $n$ on 4 solutions for different values of $\nu$. Here $J = 1$ and $T = 0.1$. 
		(a) $\nu = 0$. (b) $\nu = \pi T \simeq 0.314$. (c) $\nu = 2$.}
	\label{fig:q=2-cartoon-sym}
\end{figure}
For every Matsubara mode, we have 4 different branches of solutions, given by the equations (\ref{G01-nu})-(\ref{G14-nu}). Every solution gives a contribution to the action density, which we denote as 
\bea
\rho^{(j)}(\omega_n, J, \nu)&=&-\fl^{(j)}+\frac{J^2}2\fm^{(j)},\,\,\,\,j=1,2,3,4
\label{action1-4-nu}\eea
We present the plots of $\rho^{(j)}$ as function of $n$ for different values of the coupling $\nu$ on the Fig.\ref{fig:q=2-cartoon-sym}. For the decoupled case $\nu = 0$, as was shown in \cite{AKTV}, the solution 1, which corresponds to the standard replica-diagonal saddle, has the lowest action, and thus lowest action density for every Matsubara mode, see Fig.\ref{fig:q=2-cartoon-sym}A. The 2nd solution, which is also replica-diagonal, has the highest action. Meanwhile, the 3rd and 4th solutions are have the same action. When $\nu$ increases, the action density of the solution 3 increases, whereas the \textit{action density of the solution 4 decreases}. As shown on the Fig.\ref{fig:q=2-cartoon-sym}B, at certain value of $\nu$ we have $\rho^{(1)}(\pm\omega_0) = \rho^{(4)}(\pm\omega_0)$. This signifies the first phase transition: after that point the solution with $G(\pm\omega_0) = G^{(4)}(\pm\omega_0)$ becomes dominant over the solution which involves the $G^{(1)}$ root for all Matsubara modes. The value of action on the dominant saddle point develops a discontinuity of the first derivative in temperature, which indicates the first order phase transition. As $\nu$ increases further, the cyan curve goes down further, and more phase transitions occur. On Fig.\ref{fig:q=2-cartoon-sym} the saddle point, which dominates, has  $G(\pm\omega_0) = G^{(4)}(\pm\omega_0)$ and $G(\pm\omega_1) = G^{(4)}(\pm\omega_1)$ (the rest of Matsubara modes are still defined by $G^{(1)}$. 

To summarize, when increasing the coupling or, equivalently, lowering the temperature, we encounter infinite number of phase transitions. Every phase transition consecutively replaces the root $G^{(1)}$ with $G^{(4)}$ for the Matsubara modes starting from $\pm\omega_0$ on the dominant saddle point. 
\begin{figure}[t]
	\centering
	\includegraphics[scale=0.58]{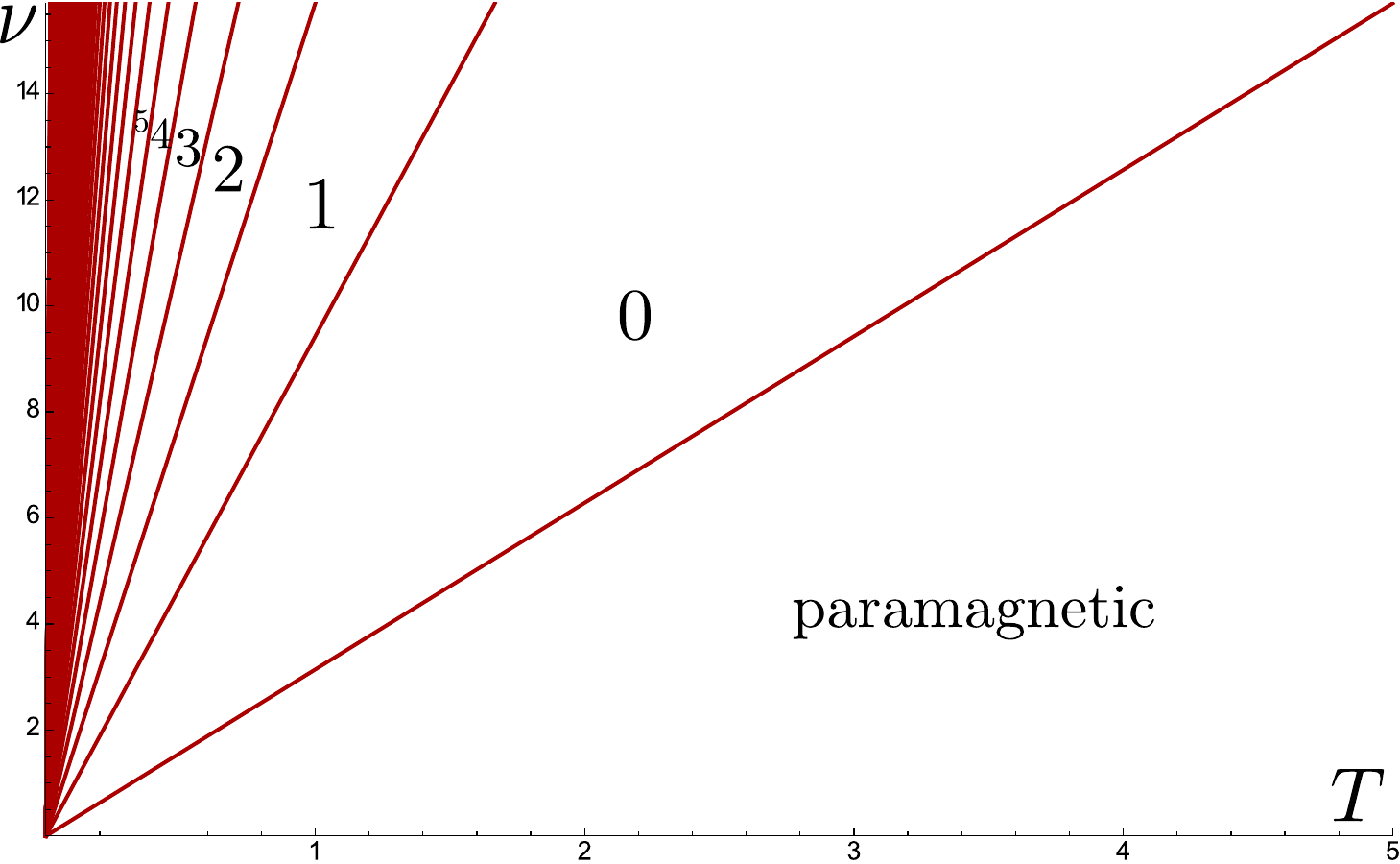}$\,\,\,$
	\caption{Phase diagram of two nonlocally coupled SYK$_2$ chains in the $\nu$-$T$ plane, as defined by the equation (\ref{Tcr-sym}). The numbers label the nontrivial phases.}
	\label{fig:q=2-PD-sym}
\end{figure}

Let us introduce some terminology for these phases. 
\begin{itemize}
	\item The phase, when the dominant saddle point has $G(\omega_n) = G^{(1)}(\omega_n) $ for all $n$, is called the \textit{paramagnetic} phase. 
	\item We will label the given non-paramagnetic phase by the bold integer number ${\bf n}_0$, if we have $G(\pm\omega_n) = G^{(4)}(\pm\omega_n) $ for $n < {\bf n}_0$, and $G(\omega_n) = G^{(1)}(\omega_n)$ for the rest of Matsubara modes. Example: phase ${\bf 0}$ is defined by  $G(\pm\omega_0) = G^{(4)}(\pm\omega_0) $ and $G(\omega_n) = G^{(1)}(\omega_n)$ for the rest. 
\end{itemize}
Consider the phase transition from the phase ${\bf n-1}$ to the phase ${\bf n}$\footnote{The paramagnetic phase here corresponds to ${\bf -1}$.}. We can find the critical curve for such phase transition from the equation
\be
\rho^{(4)}(\omega_n(T), J, \nu) = \rho^{(1)}(\omega_n(T), J, \nu)
\ee
for any $n$. Doing some straightforward but tedious algebra, one can check that this equation is solved by the linear law: 
\be
\nu_{cr} = 2\pi T_{cr} \left(n + \frac12\right)\,. \label{Tcr-sym} 
\ee
Note that there is no dependence on $J$. The corresponding phase diagram in the plane of replica coupling $\nu$ and temperature $T$ is presented on the Fig.\ref{fig:q=2-PD-sym}. Let us make some comments on it: 
\begin{itemize}
	\item At any given temperature (coupling), one can probe infinite number of phases by increasing the coupling (decreasing the temperature). 
	\item At any given temperature, when decreasing the coupling quasi-statically, one always ends up in the paramagnetic phase. 
\end{itemize}

\section{Quartic case}
\label{sec:q=4}
\subsection{Numerical solution of saddle point equations}

Having studied the analytically tractable $q=2$ case, we now proceed to the study of saddle points in the interacting $q=4$ case. Specifically, we numerically solve the saddle point equations (\ref{saddle-point-1-eta})-(\ref{saddle-point-2-eta}) with the source (\ref{eta_sym}) numerically, using the replica-symmetric ansatz (\ref{NDS}). Under these assumptions, the saddle point equations read
\bea
&& -i \omega_n G_0(\omega_n) - G_0(\omega_n) \Sigma_0(\omega_n) - G_1(\omega_n) \Sigma_1(\omega_n) =1\,;\label{SD-RS-1-diag}\\
&& i \omega_n G_1(\omega_n) + G_1(\omega_n) \Sigma_0(\omega_n)+ G_0(\omega_n) \Sigma_1(\omega_n)=0\,;\label{SD-RS-1-offdiag}\\
&& \Sigma_{0}(\tau, \tau') = J^2 G_{0}(\tau, \tau')^{q-1}\,;\label{SD-RS-2-diag}\\
&& \Sigma_{1}(\tau, \tau') = J^2 G_{1}(\tau, \tau')^{q-1} + \zeta(\tau-\tau')\,.\label{SD-RS-2-offdiag}
\eea
We solve the equations at finite temperature, so that $\zeta$ is given by (\ref{eta_sym(t)T}) in the position space. To construct the numerical solutions, we use the approach described in section 4.1 of \cite{AKTV}, generalized to the case of interacting replicas. The fixed parameters of the system are $J$, $\nu$ and the temperature $T = \beta^{-1}$. The initial condition for the iteration procedure is given by a solution of the $q=2$ version of the model. The resulting solution is obtained by iterating the equations of motion and waiting for convergence until the desired accuracy. 
\begin{figure}[t!]
	\centering
	\includegraphics[scale=0.4]{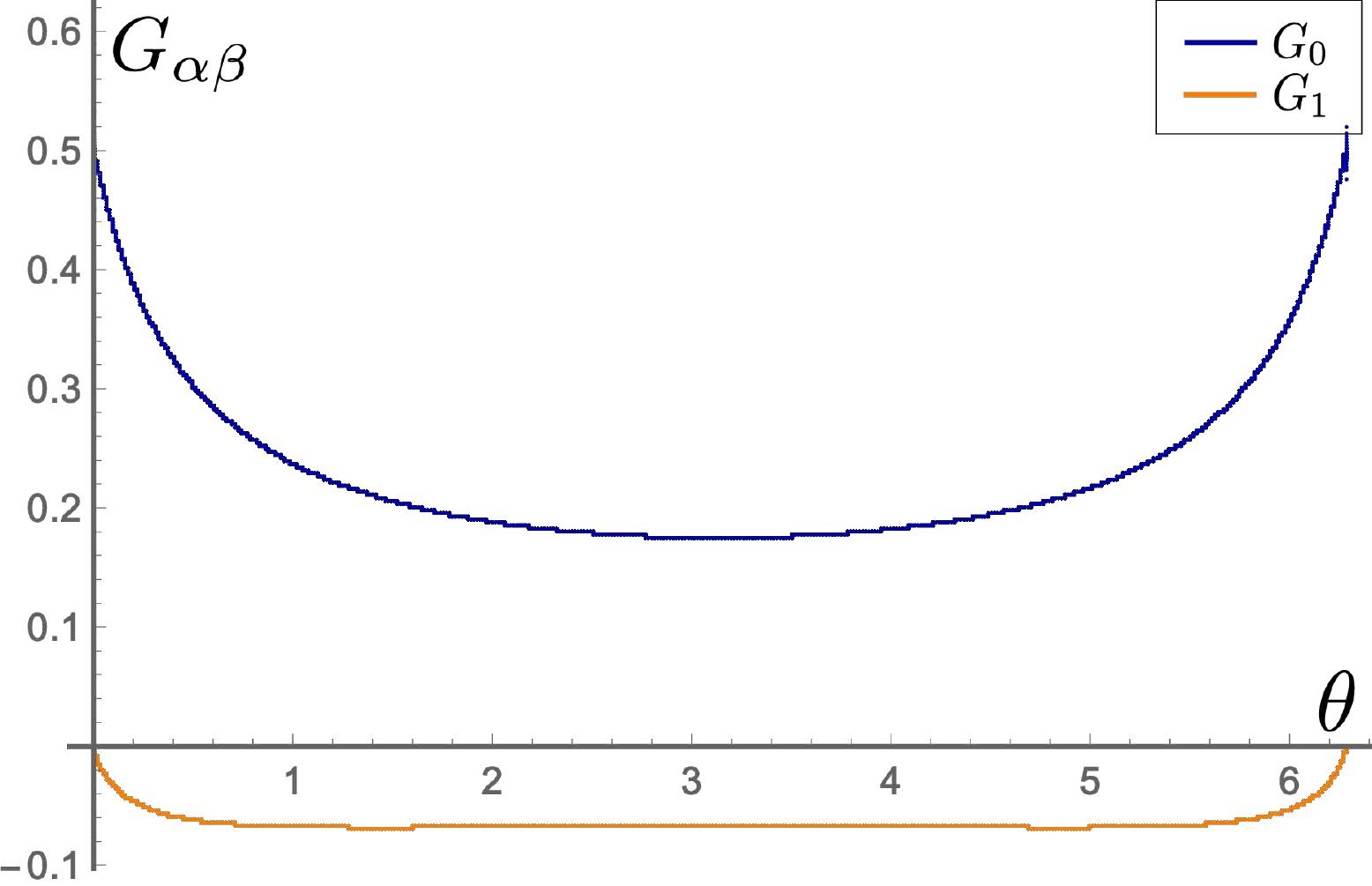}
	\includegraphics[scale=0.4]{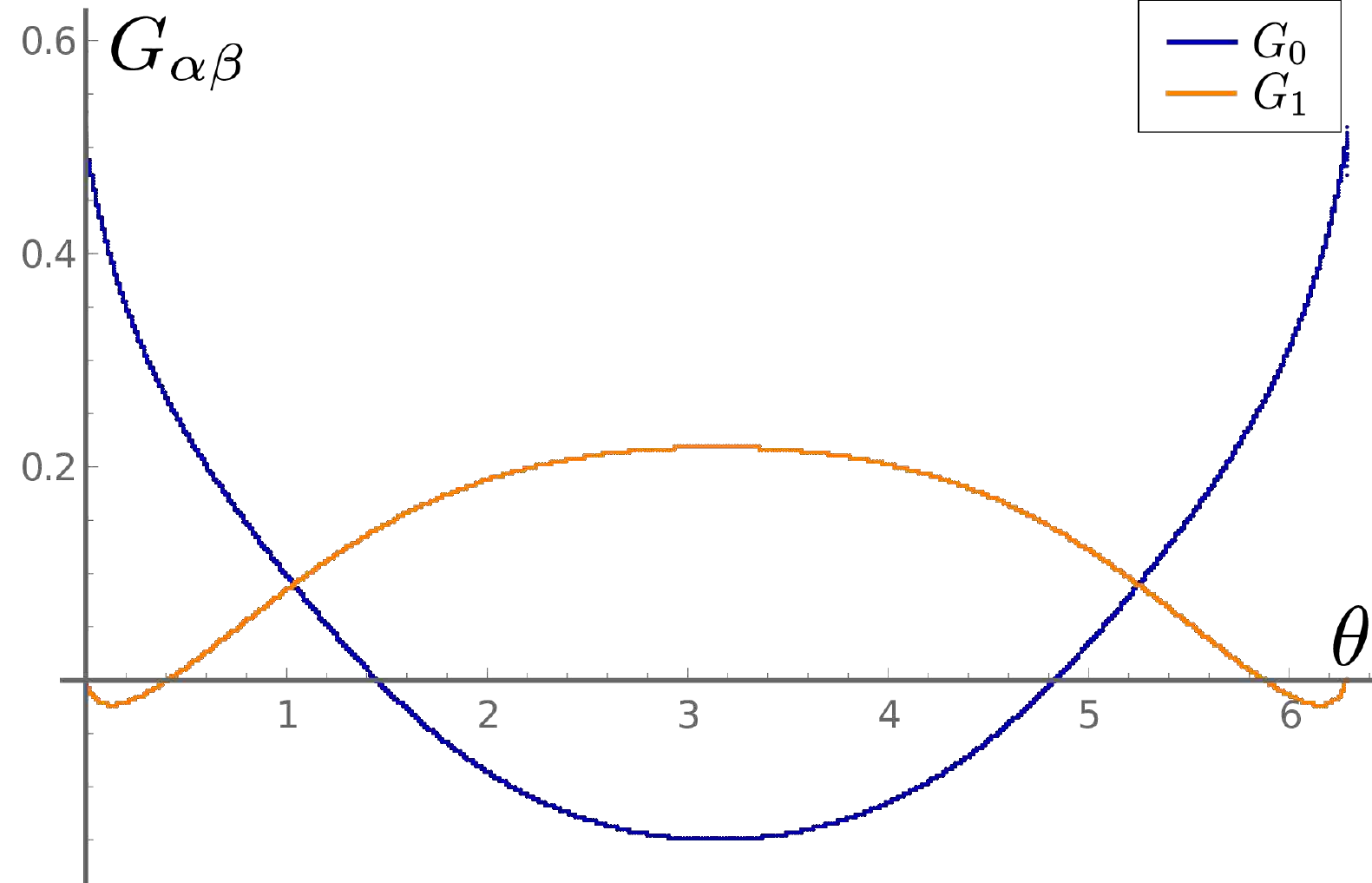}\\
	(a) $\qquad \qquad \qquad \qquad \qquad \qquad \qquad $ (b)\\
	\includegraphics[scale=0.4]{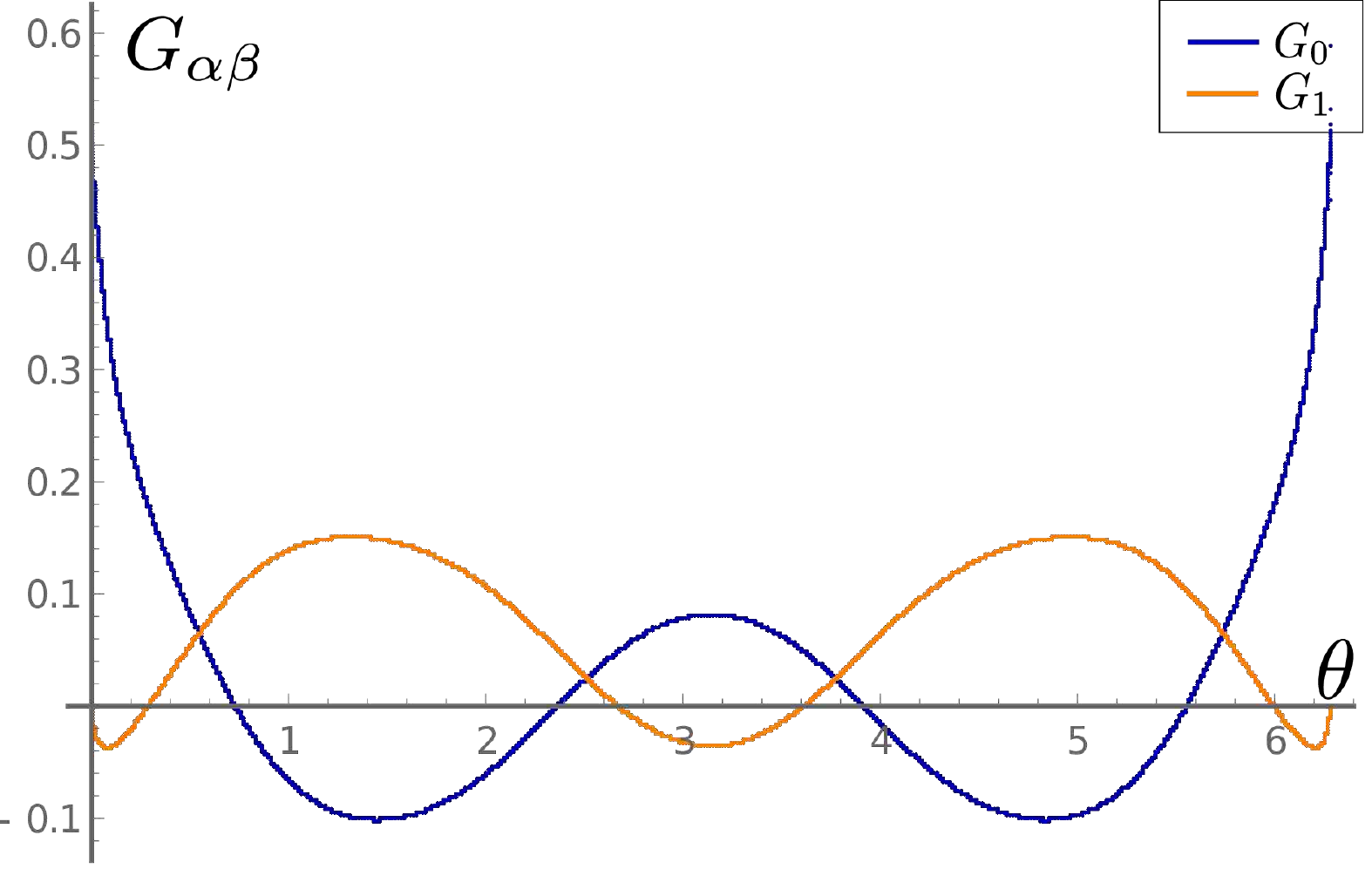}
	\includegraphics[scale=0.4]{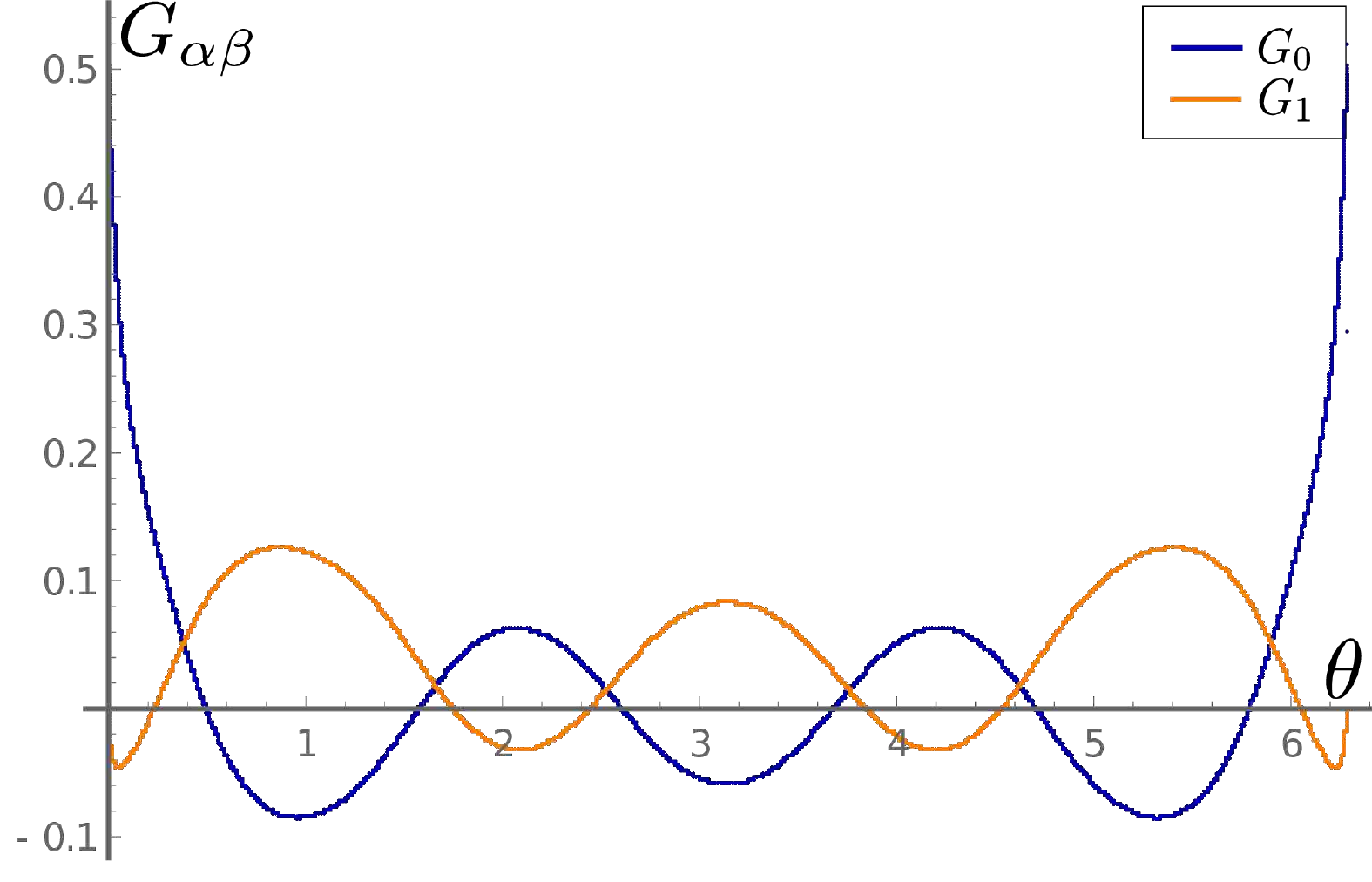}\\
	(c) $\qquad \qquad \qquad \qquad \qquad \qquad \qquad $ (d)\\
	\caption{Numerical solutions for $G_{\alpha\beta}$ as function of $\theta = \frac{2\pi}{\beta} \tau$ for $\nu = 0.1$ and $J=1$. (a) Paramagnetic phase, $T=0.032$. (b) Phase \textbf{0}, $T=0.025$. (c) Phase \textbf{1}, $T=0.008$. (d) Phase \textbf{2}, $T=0.004$.}
	\label{fig:sol-nu=01}
\end{figure}
\newpage
\begin{itemize}
	\item Like in the decoupled replicas case \cite{AKTV}, for every $q=2$ solution there exists a solution in the $q=4$ model.
	\item It is observed that for solutions that we studied, the hierarchy in the dominance of the saddle points is very similar to the $q=2$ case. Different phases correspond to solutions, which are obtained by iterating from the dominant solutions in the $q=2$ model. In other words, the competition happens between the solutions, which are obtained from $q=2$ trial functions $G_{\alpha\beta}(\omega_n)$ containing different number of Matsubara modes with the root 4 (\ref{G04-nu})-(\ref{G14-nu}). 
	\item We use the same terminology for phases as in the $q=2$ case in section \ref{sec:phases-q=2}. 
\end{itemize}
\begin{figure}[t!]
	\centering
	\includegraphics[scale=0.4]{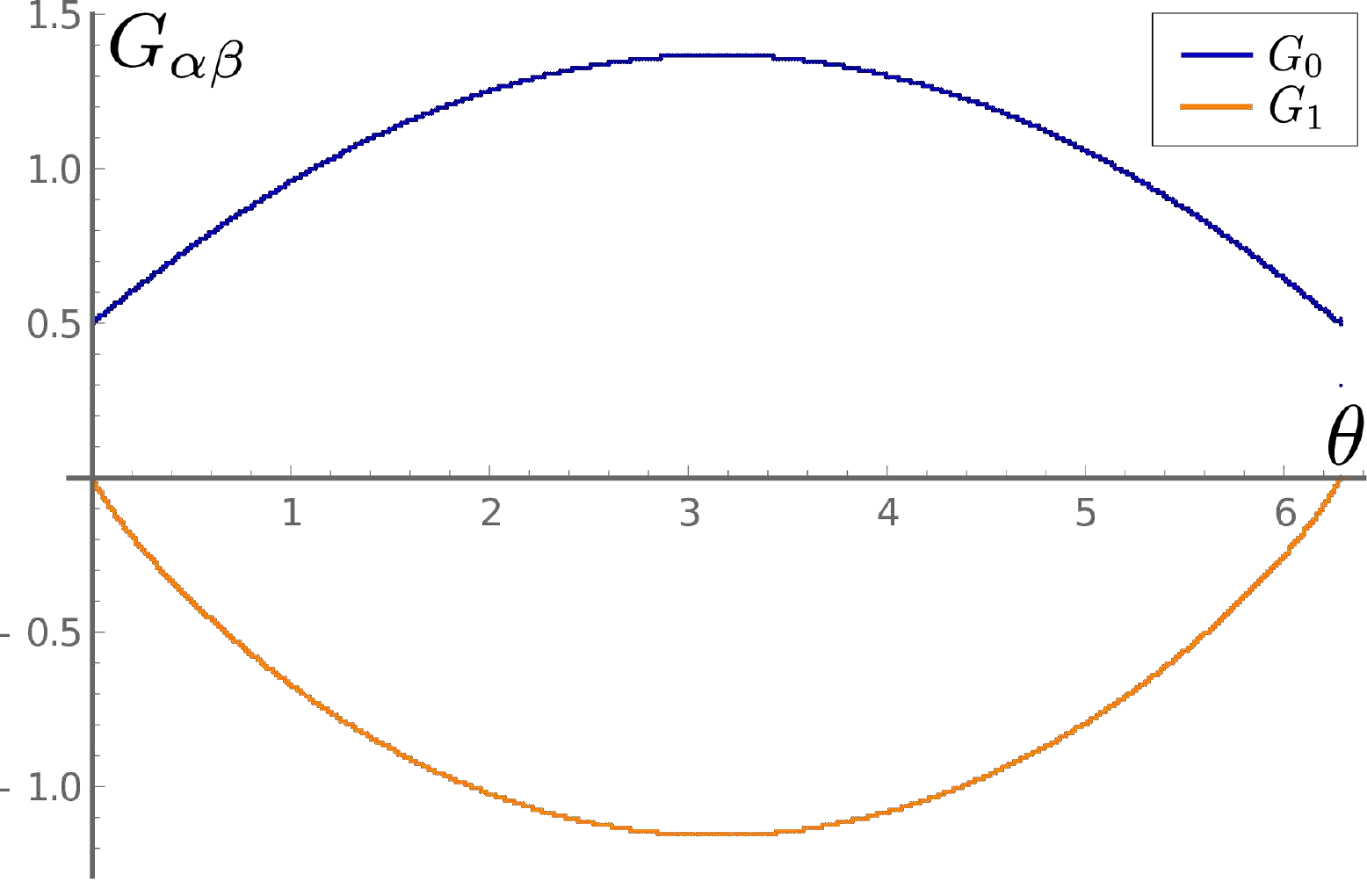}
	\includegraphics[scale=0.4]{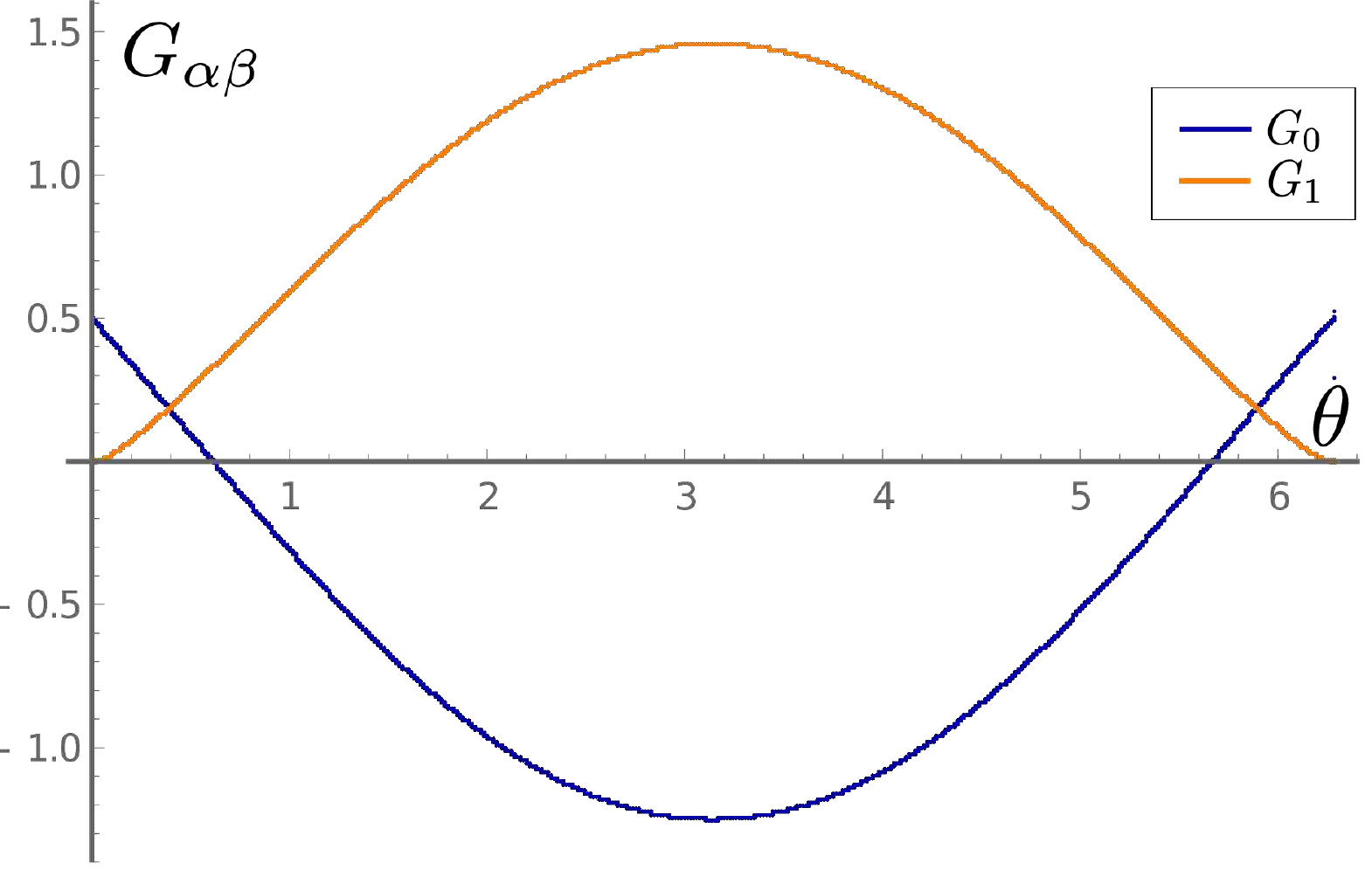}\\
		(a) $\qquad \qquad \qquad \qquad \qquad \qquad \qquad $ (b)\\
	\includegraphics[scale=0.4]{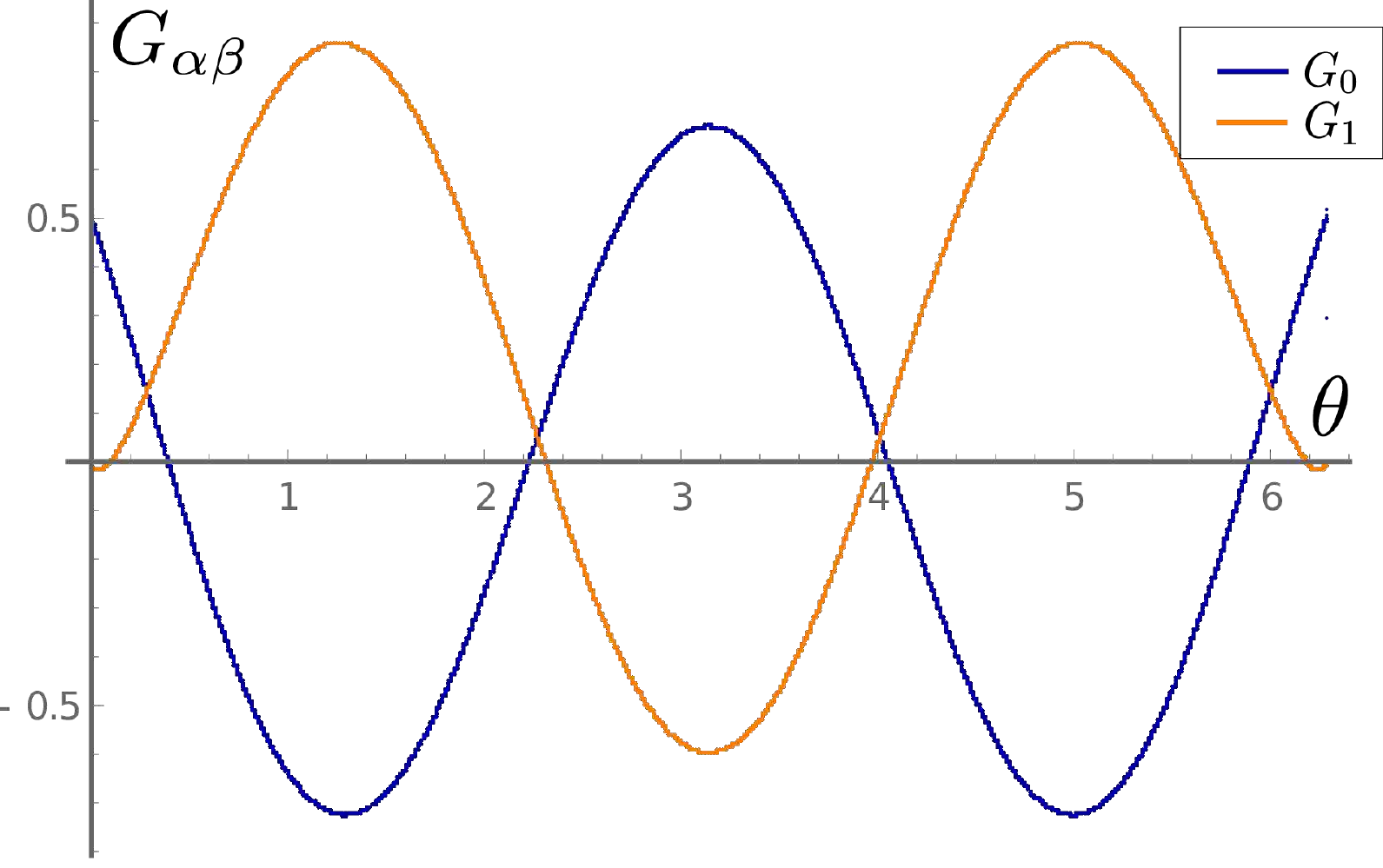}
	\includegraphics[scale=0.4]{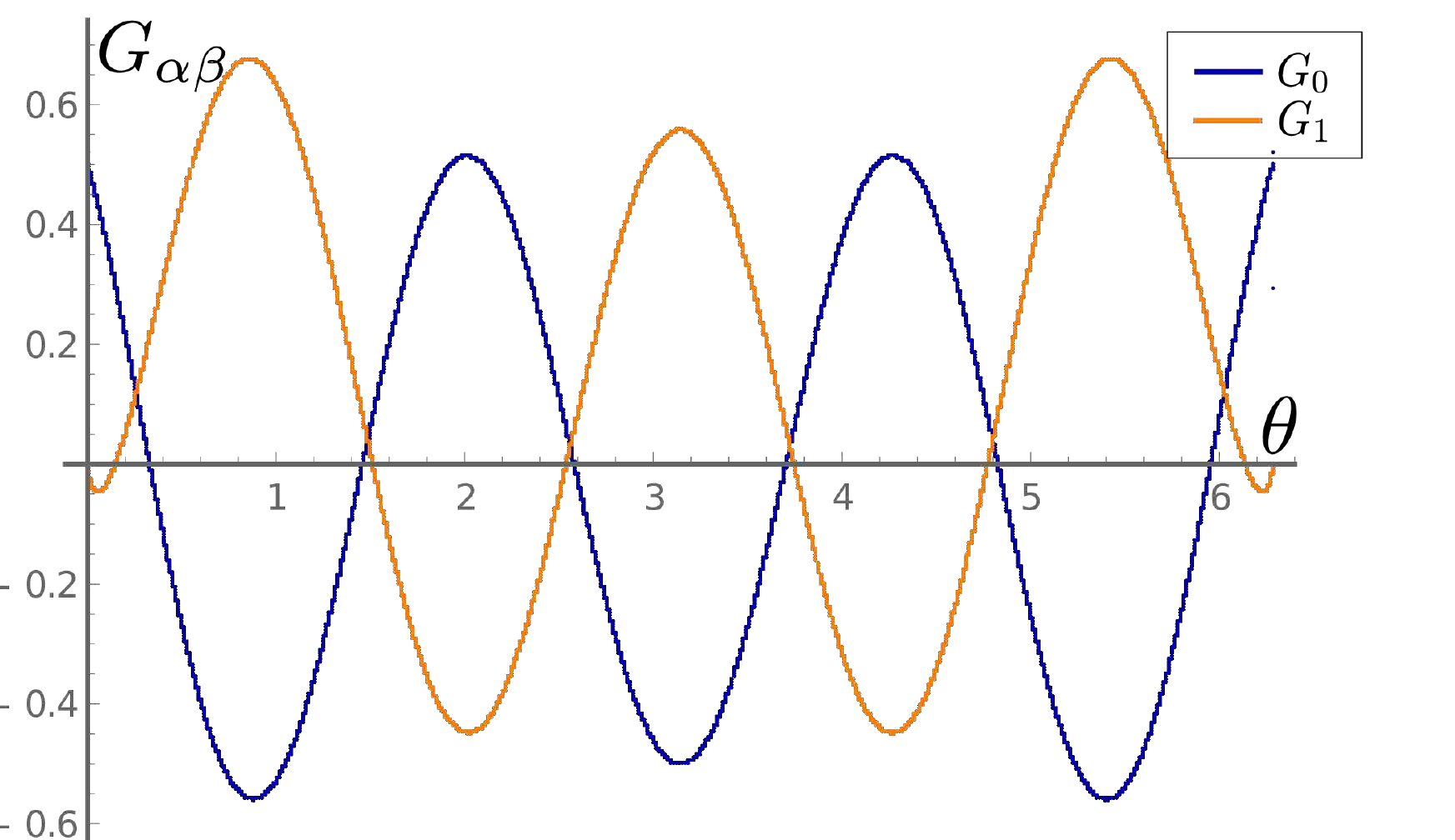}\\
		(c) $\qquad \qquad \qquad \qquad \qquad \qquad \qquad $ (d)\\
	\caption{Numerical solutions for $G_{\alpha\beta}$ as function of $\theta = \frac{2\pi}{\beta} \tau$ for $\nu = 5$ and $J=1$. (a) Paramagnetic phase, $T=1.6$. (b) Phase \textbf{0}, $T=1.55$. (c) Phase \textbf{1}, $T=0.51$. (d) Phase \textbf{2}, $T=0.3$.}
	\label{fig:sol-nu=5}
\end{figure}

\subsection{Phase structure}
\label{sec:q=4-phases}

To study the phase structure of the model, we compute the action (\ref{I(eta)}) on numerical solutions of equations (\ref{SD-RS-1-diag})-(\ref{SD-RS-2-offdiag}). The general expression for the action reads
\bea
 I[G, \Sigma; \eta] &=& - \log \Pf[\delta_{\alpha\beta} \dd_\tau - \hat{\Sigma}_{\alpha\beta}] +\frac12 \int_0^\beta \int_0^\beta d\tau_1 d \tau_2  \left(\Sigma_{\alpha\beta}(\tau_1, \tau_2) G_{\alpha\beta}(\tau_1, \tau_2) - \frac{J^2}{q} G_{\alpha\beta}(\tau_1, \tau_2)^q\right) \nn\\&&-\frac{1}{2} \int_0^\beta \int_0^\beta d \tau_1 d\tau_2 \eta_{\alpha\beta}(\tau_1, \tau_2) G_{\alpha\beta}(\tau_1, \tau_2)\,, \label{Ion-shell}
\eea
We use equations of motion, subtract the free part from the logarithm and assume that fields depend only on $u = \tau_1 - \tau_2$. Then one gets 
\bea
I|_{\text{on-shell}} &=& - \frac12\sum_n \log \det \left[\delta_{\alpha\beta} + \frac{\Sigma_{\alpha\beta}(\omega_n)}{i\omega_n}\right]  + \left(1-\frac1q \right)J^2 \int_0^\beta d u \sum_{\alpha,\beta} G_{\alpha\beta}(u)^{q} \Big|_{\text{on-shell}}\,. \label{Ion-shell2}
\eea

Constructing the numerical solutions of saddle point equations and computing the on-shell action, we find that the phase structure of the $q=4$ model is similar to the $q=2$ model: when decreasing the temperature at fixed coupling, the system goes through infinite number of phase transitions. The solutions which correspond to different phases are presented on Fig.\ref{fig:sol-nu=01} and Fig.\ref{fig:sol-nu=5}. 
\begin{itemize}
	\item[1.] On Fig.\ref{fig:sol-nu=01} the coupling between replicas is weak, compared to the SYK self-interaction of fermions, $\nu = 0.1 < J=1$. The behavior of solutions in non-paramagnetic phases have many similarities with the replica-nondiagonal solutions of the decoupled replicas from \cite{AKTV}. The effect of the coupling $\nu$ comes down to shifting the long-time region by a constant and introducing extra local extrema in the short time region. 
	\item[2.] On Fig.\ref{fig:sol-nu=5} the coupling between replicas is strong, compared to the SYK self-interaction of fermions, $\nu = 5 > J=1$. In this case the oscillatory behavior dominates over the regular SYK$_4$-like dynamics.
\end{itemize}

\begin{figure}[t!]
	\centering
	\includegraphics[scale=0.5]{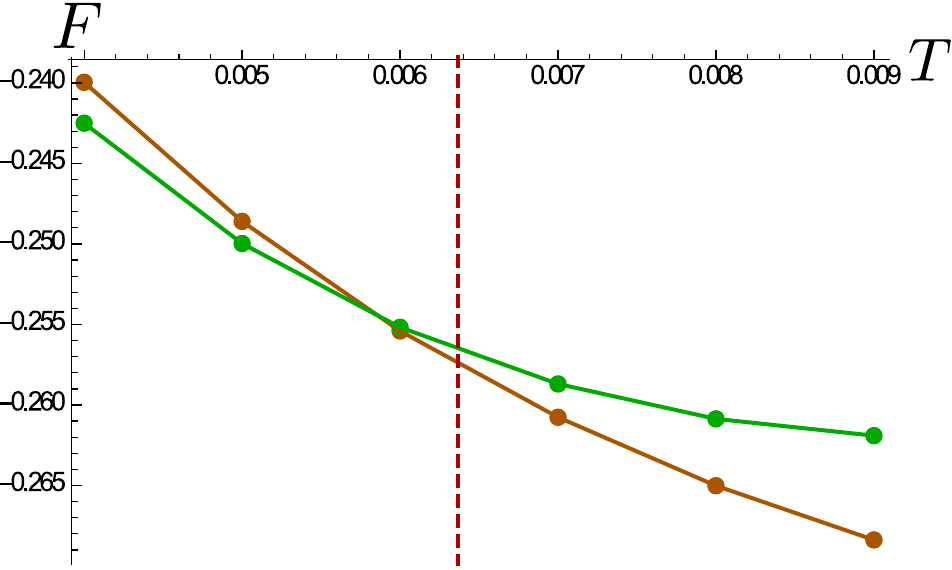}
	\includegraphics[scale=0.5]{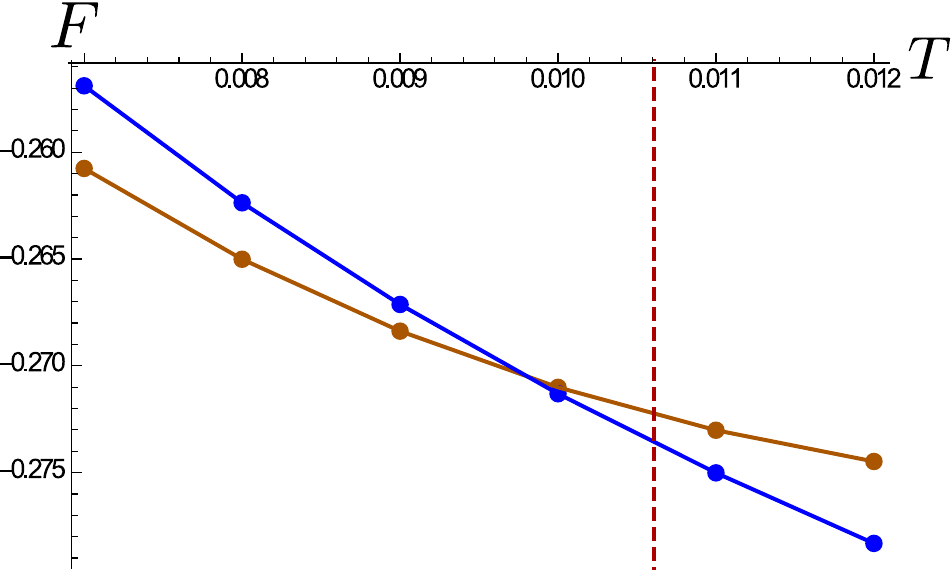}
	\includegraphics[scale=0.5]{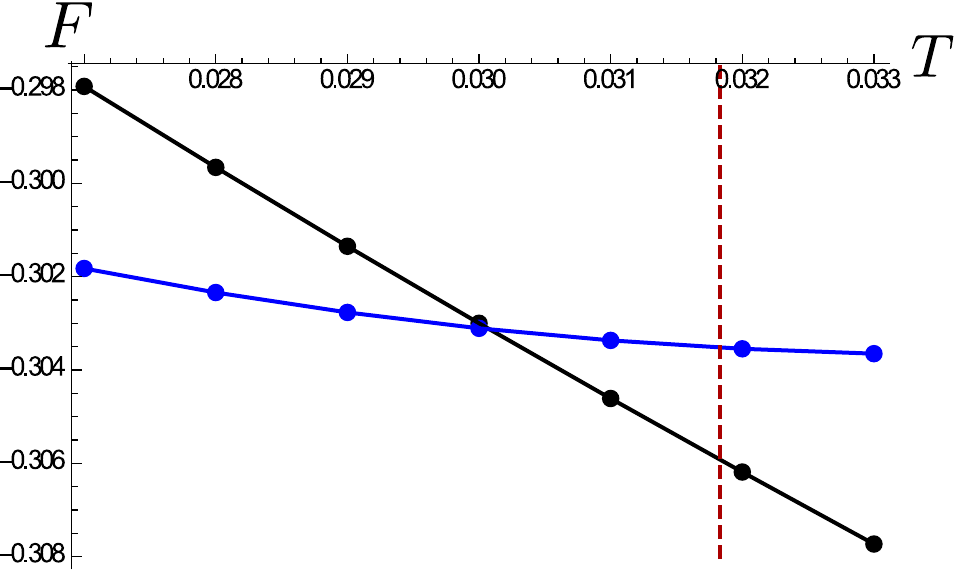} \\
	 (a) $\qquad \qquad \qquad \qquad \qquad$ (b) $\qquad \qquad \qquad \qquad \qquad$ (c)\\
	\includegraphics[scale=0.5]{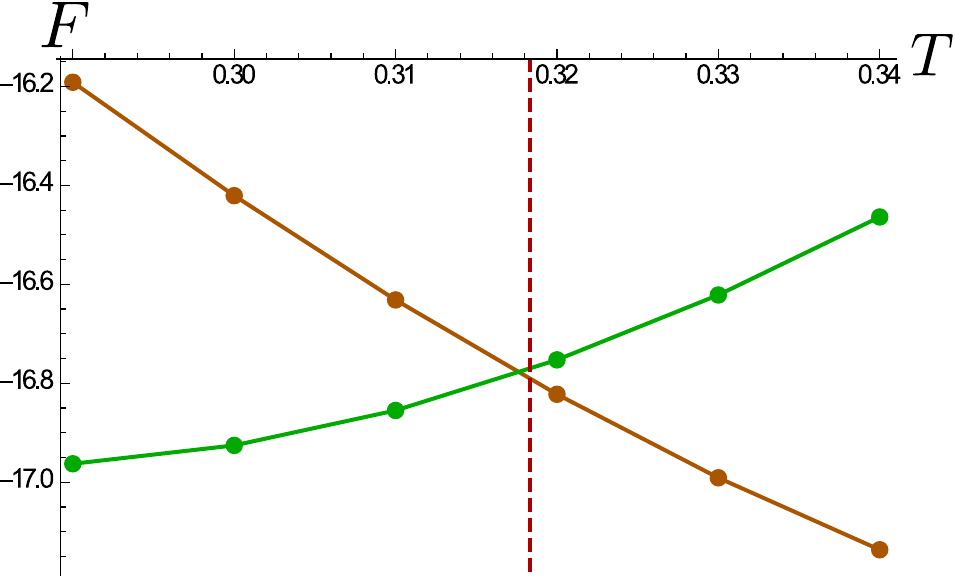}
	\includegraphics[scale=0.5]{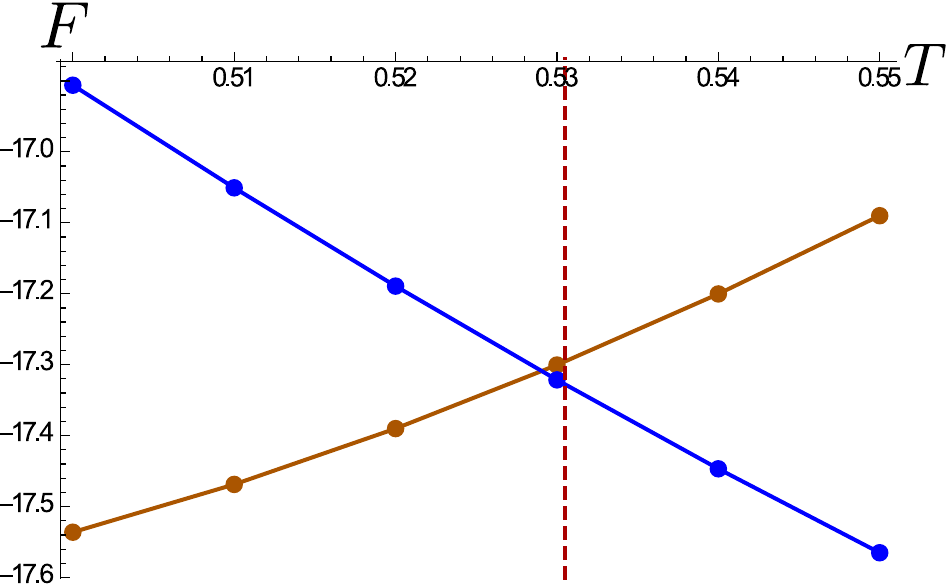}
	\includegraphics[scale=0.51]{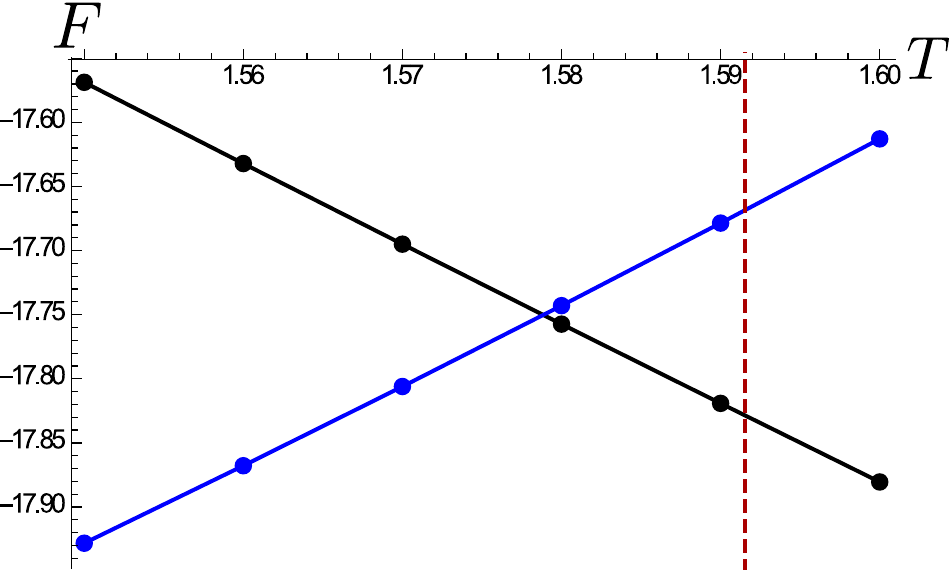}\\
	 (d) $\qquad \qquad \qquad \qquad \qquad$ (e) $\qquad \qquad \qquad \qquad \qquad$ (f)
	\caption{Free energy during phase transitions in the $q=4$ model. Here black curve is the free energy of the paramagnetic phase, blue curve is the free energy of the phase \textbf{0}, brown curve is the free energy of the phase \textbf{1} and green curve is the free energy of the phase \textbf{2}. Plots (a)-(c) correspond to $\nu = 0.1$, and plots (d)-(f) correspond to $\nu = 5$. For all plots $J = 1$. Dashed line shows for comparison the corresponding critical point of the $q=2$ model, as dictated by equation (\ref{Tcr-sym}).}
	\label{fig:q=4-phase-transitions}
\end{figure}

To study the phase transitions, we define the annealed free energy as 
\be
F = -T \log \mZ(\eta)\,,\label{F}
\ee
where $T = \beta^{-1}$. On the dominant saddle point in the leading order of large $N$, we express it through the on-shell action by 
\be
F = T N \left(-\log 2 + I|_{\text{on-shell}}\right)\,. \label{F=S/b}
\ee
Here the $\log 2$ is subtracted in order for the entropy to match the free result $N \log 2$ for $J = \nu = 0$. 

We plot the free energy near first three critical points on Fig.\ref{fig:q=4-phase-transitions}. Specifically the following phase transitions are demonstrated (in order of the increasing temperature): 
\begin{itemize}
	\item Fig.\ref{fig:q=4-phase-transitions}(a),(d): phase \textbf{1} $\rightarrow$ phase \textbf{2};
	\item Fig.\ref{fig:q=4-phase-transitions}(b),(e): phase \textbf{0} $\rightarrow$ phase \textbf{1};
	\item Fig.\ref{fig:q=4-phase-transitions}(c),(f): paramagnetic phase $\rightarrow$ phase \textbf{0};
\end{itemize}
First of all, we confirm that these phase transitions are of the first order. Second, we see that, while the paramagnetic phase has approximately linear behavior of the free energy, which is characteristic for SYK models \cite{MScomments,Maldacena18,Kim19,Garcia-Garcia19}, the nontrivial phases \textbf{0}, \textbf{1}, \textbf{2}, $\dots$ have largely nonlinear behavior of the free energy. When the coupling $\nu$ is weak (see Fig.\ref{fig:sol-nu=01}), the total free energy is monotonically decreasing, as shown on Fig.\ref{fig:q=4-phase-transitions}(a)-(c). However, when the coupling between replicas is strong (see \ref{fig:sol-nu=5}), the free energy is non-monotonic, and increases towards every critical point, as shown on Fig.\ref{fig:q=4-phase-transitions}(d)-(f). 

One point that seems common for both cases of strong and weak replica interaction is that the entropy is decreasing with temperature on the nontrivial phases \textbf{0}, \textbf{1}, \textbf{2}, $\dots$. That means that the heat capacity $C = -T \frac{\dd^2 F}{\dd T^2}$ is negative for these phases. Moreover, in the case $\nu = 5$ the growth of the free energy indicates negative entropy. This might hint that one perhaps should restrict the allowed values of $\nu$ in such a way that the entropy remains positive. Note, however, that the heat capacity on nontrivial phases is negative for all values of $\nu$. 

For comparison, we have also shown on Fig.\ref{fig:q=4-phase-transitions} the critical points of the $q=2$ model, defined by the phase diagram equation (\ref{Tcr-sym}) on Fig.\ref{fig:q=4-phase-transitions} by the dashed line. This demonstrates that the effect of the $q=4$ self-interaction is that the critical temperatures decrease compared to the $q=2$ case, and the difference is largest for the highest temperature critical point of the transition between the paramagnetic and 0th phases.

\section{Local replica coupling}
\label{sec:Wormholes}

We have also conducted analogous study of two coupled SYK models, which are related to the eternal traversable wormhole \cite{Gao16,Maldacena17,Maldacena18,Garcia-Garcia19}. In terms of the bilocal replica field action (\ref{I(eta)}), this model is obtained, when one assumes that the source $\eta_{\alpha\beta}$ has the replica-antisymmetric form
\be
\hat{\eta}(\tau_1-\tau_2) = \left(\begin{matrix}
	0 & \zeta(\tau_1-\tau_2) & \\
	- \zeta(\tau_1-\tau_2) & 0 &
\end{matrix}\right)\,,
\ee
where\footnote{Note that the sign of $\mu$ in our case is opposite to that of \cite{Maldacena18}. While the sign is significant for the gravity analysis \cite{Gao16,Maldacena18} and for exact diagonalization studies \cite{Maldacena18,Garcia-Garcia19}, it is unimportant for solution of saddle point equations.} 
\be
\zeta(\tau) =i \mu \delta(\tau)\,. \label{zeta_antisym(t)}
\ee 
In the momentum space one simply has
\be
\zeta(\omega_n) = i \mu\,. \label{zeta_antisym}
\ee
This results in the local coupling of the traversable wormhole models \cite{Maldacena17,Maldacena18} of the form 
\be
H_{\text{int}} =-i \mu \sum_i \psi_i^L (\tau) \psi_i^R (\tau)\,.
\ee
We expect that the replica-antisymmetric source supports solution for $G$, which have the form: 
\be\label{NDAS}
G_{LL}(\tau_1, \tau_2)=G_{RR}(\tau_1, \tau_2)=G_{0}(\tau_1, \tau_2),\,\,\,\,\,G_{LR} (\tau_1, \tau_2)=-G_{RL}(\tau_1, \tau_2)=G_1(\tau_1, \tau_2)\,.
\ee
The dynamic variables are $G_0$ and $G_{1}$. Taking into account the antisymmetry condition (\ref{antisymmetry}), the ansatz (\ref{NDAS}) implies that $G_0$ must be an odd function in the frequency space, whereas $G_{1}$ must be an even function\footnote{These assumption about the symmetry properties of functions are in line with the properties of correlators in thermofield double and other studies of coupled SYK models \cite{Kim19,Garcia-Garcia19,Maldacena18,Saad18}.}. 

\subsection{The case $q=2$}

Like in the case of the replica-symmetric source, one can also solve the model in the $q=2$ case. Assuming the ansatz (\ref{NDAS}) for $G$, the saddle point equations (\ref{saddle-point-1-eta})-(\ref{saddle-point-2-eta}) have the form
\bea
\Sigma_0(\omega_n) &=& J^2 G_0(\omega)\,; \\
\Sigma_{1}(\omega) = -\Sigma_{RL}(\omega)  &=& J^2 G_{1}(\omega) + i \mu \,,
\eea
and we have equations in the frequency space
\bea\label{eq1-mu}
-i\omega G_0(\omega) -J^2 G_0(\omega)^2+ (J^2 G_{1}(\omega)+i \mu) G_{1}(\omega) &=&1\,; \\
\label{eq2-mu}-i \omega G_{1}(\omega) -J^2 G_0(\omega) G_{1}(\omega) - (J^2 G_{1}(\omega) +i \mu) G_0(\omega) &=&0 \,.
\eea
Note that the antisymmetric matrices do not form a closed algebra under the matrix multiplication, unlike the symmetric matrices. Therefore, for arbitrary replica number $M$ there is no guarantee for a consistent system of equations that would allow for a nontrivial replica-nondiagonal solution. Fortunately, this is the case for $M=2$. 

\subsubsection{Solutions}

The solutions for general value of $\mu$ are given by: 
\bea
G_0^{(1)}(\omega_n) &=& -\frac{i}{2 J^2} \left[\omega_n - \frac{\sgn(\omega_n)}{\sqrt{2}}C_1\right]\,; \label{G01-mu}\\ 
G_1^{(1)}(\omega_n) &=& \frac{i}{8 J^2 \mu |\omega_n|} \left(-\sqrt{2} C_1 (4 J^2 + \omega_n^2 - D) - \mu^2 (4 |\omega_n| - \sqrt{2} C_1) \right)\,;\label{G11-mu}\\
G_0^{(2)}(\omega_n) &=& -\frac{i}{2 J^2} \left[\omega_n + \frac{\sgn(\omega_n)}{\sqrt{2}} C_1  \right]\,;\label{G02-mu}\\
G_1^{(2)}(\omega_n) &=& \frac{i}{8 J^2 \mu |\omega_n|} \left(\sqrt{2} C_1 (4 J^2 + \omega_n^2 - D) - \mu^2 (4 |\omega_n| + \sqrt{2} C_1) \right)\,;\label{G12-mu}\\
G_0^{(3)}(\omega_n) &=& -\frac{i}{2 J^2} \left[\omega_n + \frac{\sgn(\omega_n)}{\sqrt{2}}C_2\right]\,;\label{G03-mu}\\
G_1^{(3)}(\omega_n) &=& \frac{i}{8 J^2 \mu |\omega_n|} \left(\sqrt{2} C_2 (4 J^2 + \omega_n^2 + D) - \mu^2 (4 |\omega_n| + \sqrt{2} C_2) \right)\,;\label{G13-mu}\\
G_0^{(4)}(\omega_n) &=& -\frac{i}{2 J^2} \left[\omega_n - \frac{\sgn(\omega_n)}{\sqrt{2}}C_2\right]\,;\label{G04-mu}\\
G_1^{(4)}(\omega_n) &=& \frac{i}{8 J^2 \mu |\omega_n|} \left(-\sqrt{2} C_2 (4 J^2 + \omega_n^2 + D) - \mu^2 (4 |\omega_n| - \sqrt{2} C_2) \right)\,,\label{G14-mu}\\
\eea
where we introduce the auxiliary notations: 
\bea
C_1 &=& \sqrt{4 J^2 + \omega_n^2 - \mu^2 + D}\,;\\
C_2 &=& \sqrt{4 J^2 + \omega_n^2 - \mu^2 - D}\,;\\
D &=& \sqrt{16 J^2 \omega_n^2 + (- 4 J^2 + \mu^2 + \omega_n^2)^2}\,.
\eea
Analogously to the analysis of section \ref{sec:q=2-solutions-sym}, one can also expand these solutions in powers of $\mu$ and see that the 3rd and 4th solutions have non-zero limit for $G_1$ as $\mu \to 0$. 

\subsubsection{Phase structure}

To study the dominance of the saddle points, we again study the action density $\rho$, defined according to (\ref{Ss}). Analogously to equations (\ref{sj})-(\ref{fm}), we again write
\bea
\rho&=&-\frac12\fl+\frac{J^2}2\,\fm\label{sj2}\,;\\
\fl&=& \log\left[\left(1+\frac{J^2 G_0(\omega_n)}{i\omega_n}\right)^2 + \left(\frac{J^2 G_1(\omega_n) + i\mu}{i\omega_n}\right)^2\right]\,;\\
\fm &=&  |G_0(\omega_n)|^2
+|G_1(\omega_n)|^2\,.
\eea
\begin{figure}[t]
	\centering
	\includegraphics[scale=0.28]{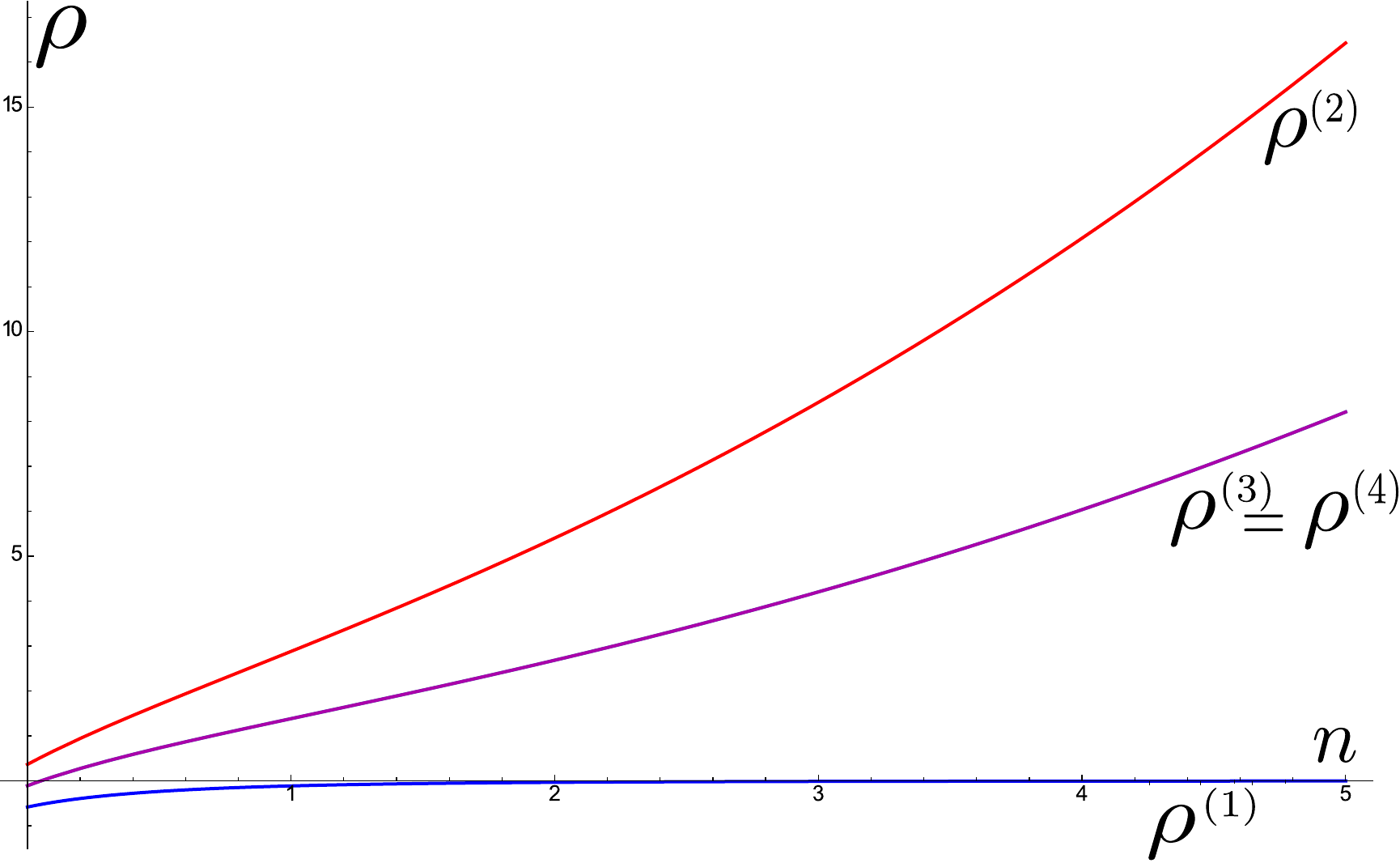}
	\includegraphics[scale=0.28]{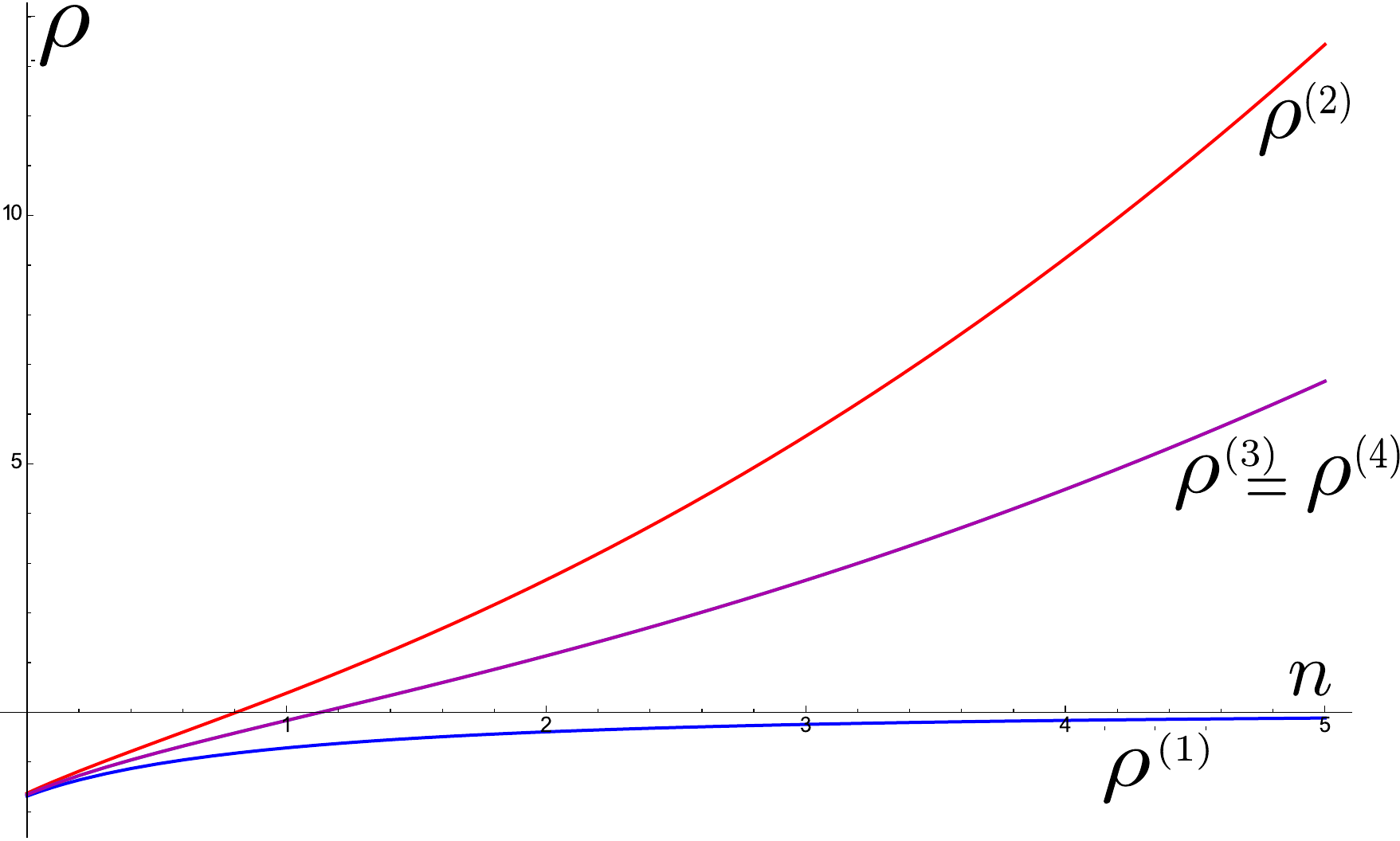} 
	\includegraphics[scale=0.28]{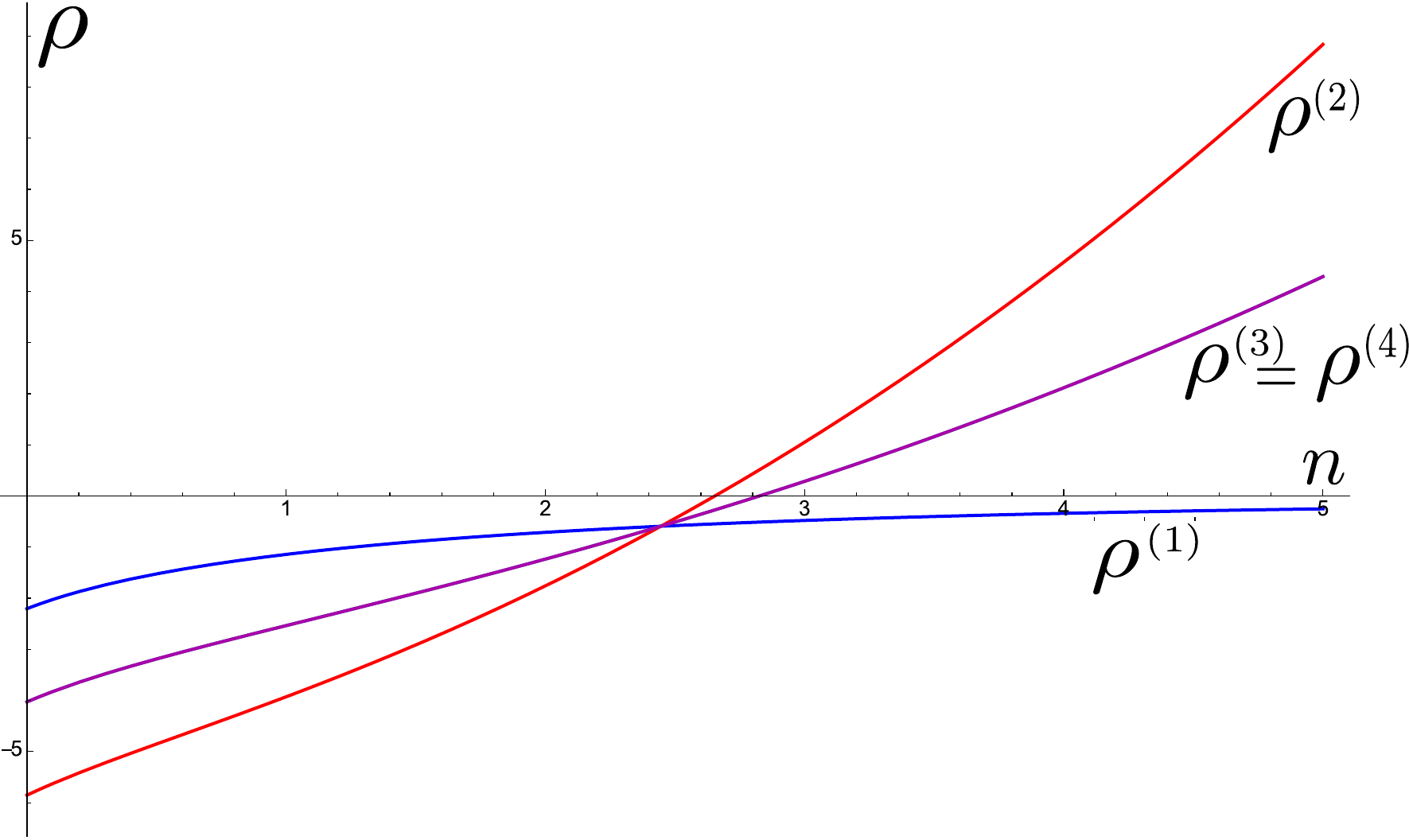}\\
	(a) $\qquad \qquad \qquad \qquad \qquad$ (b) $\qquad \qquad \qquad \qquad \qquad$ (c)\\
	\caption{Action densities \eqref{action1-4-nu} as functions of Matsubara frequency label $n$ on 4 roots for different values of $\mu$. Here $J = 1$ and $T = 0.1$. 
		(a) $\mu = 0$. (b) $\mu = 1.76$. (c) $\nu = 2.8$.}
	\label{fig:q=2-cartoon-antisym}
\end{figure}
We plot the action densities $\rho^{(i)}(\omega_n, J, \mu)$ for the four solutions as functions of $n$ on Fig.\ref{fig:q=2-cartoon-antisym}. We see that the phase structure of the model with local replica-antisymmetric coupling is very different from the model with the nonlocal replica-symmetric coupling (see Fig.\ref{fig:q=2-cartoon-sym}). In the present case, the competition now happens between the solution 1 and the solution 2. On the plot Fig.\ref{fig:q=2-cartoon-antisym}(a), corresponding to the decoupled case $\mu = 0$, the solution 1 dominates. On the plot Fig.\ref{fig:q=2-cartoon-antisym}(b) the critical point is shown, when $\rho^{(2)}(\omega_0, J, \mu) = \rho^{(1)}(\omega_0, J, \mu)$. Finally, on the plot Fig.\ref{fig:q=2-cartoon-antisym}(c) the dominant solution is the one which has $G_{\alpha\beta}(\pm\omega_k)=G^{(2)}_{\alpha\beta}(\pm\omega_k)$ for $k=0,1,2$ and $G_{\alpha\beta}(\pm\omega_n)=G^{(1)}_{\alpha\beta}(\pm\omega_n)$ for the rest of Matsubara modes. Also, note that the coupling $\mu$ does not lift the degeneracy between solutions 3 and 4, unlike in the nonlocal case. The saddle points with solutions 3 and 4 are always subleading. 
\begin{figure}[t]
	\centering
	\includegraphics[scale=0.8]{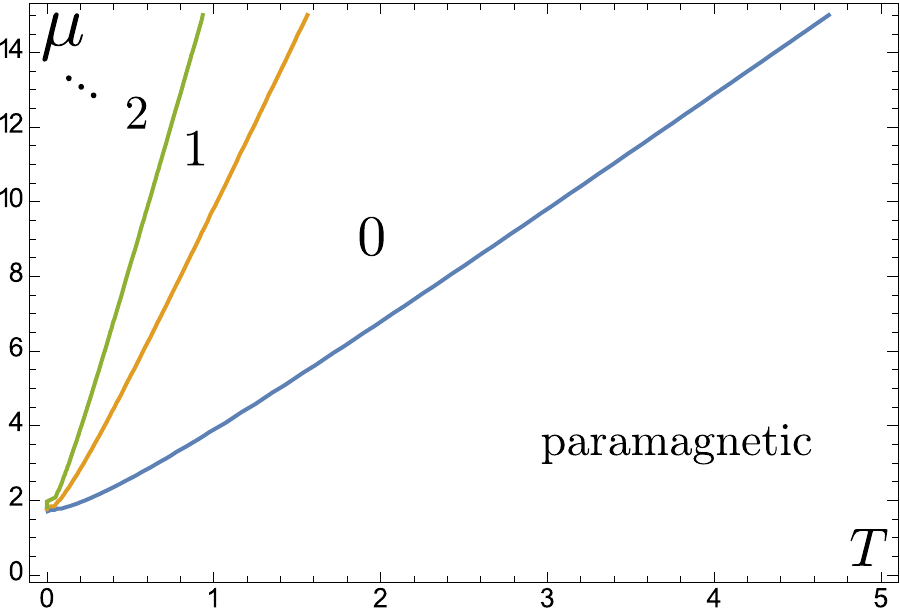}
	\caption{Phase diagram for replica-antisymmetric solutions with local coupling $\mu$ in the $\mu$-$T$ plane, as defined by the equation (\ref{phase-diagram-antisym}). The numbers label the nontrivial phases. Here $J = 1$. Only the critical curves which correspond to the three highest temperature phase transitions are shown.}
	\label{fig:q=2-PD-antisym}
\end{figure}

The critical curves are defined by the equations 
\be
\rho^{(2)}(\omega_n(T), J, \mu) = \rho^{(1)}(\omega_n(T), J, \mu) \label{phase-diagram-antisym}
\ee
for any $n$. This equation does not have an analytic solution, but it can be solved numerically to obtain the phase diagram. On the plot \ref{fig:q=2-cartoon-antisym} we plot the three highest temperature critical curves. Like in the symmetric nonlocal case, we have an infinite number of phases. The notable distinction here is the presence of gap in $\mu$: at small enough (but non-zero) coupling $\mu < \mu_0$ there are no phase transitions, and the paramagnetic phase is unique. 

\subsection{The case $q=4$: traversable wormhole dual}

The numerical study of the exact saddle point equations of the system with the source (\ref{zeta_antisym(t)}) in the case of $q=4$ was performed in \cite{Maldacena18,Garcia-Garcia19}. The main phenomenon that was discovered is the presence of the phase transition which is dual to the Hawking-Page phase transition in the bulk. We have also applied our approach of numerical integration explained in \cite{AKTV} and outlined in section \ref{sec:q=4} to this model. Let us summarize our findings: 

\begin{figure}[t]
	\centering
	\includegraphics[scale=0.5]{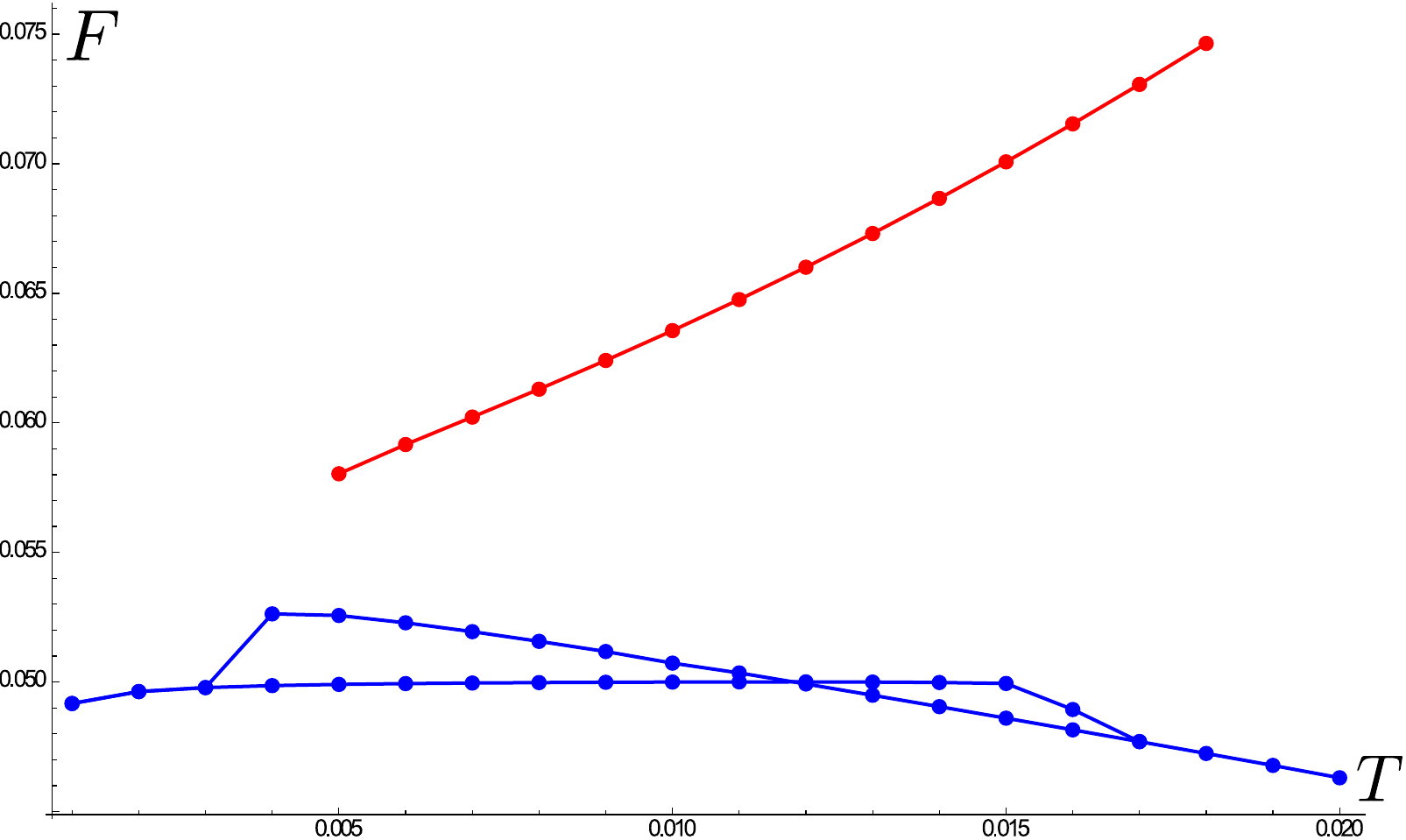}
	\caption{The Maldacena-Qi hysteresis of the free energy (blue curve), together with free energy of a subleading saddle (red curve). Here $J = \sqrt{2}$ and $\mu = 0.03$. }
	\label{fig:q=4-hysteresis}
\end{figure}
\begin{itemize}
	\item At temperatures that are significantly higher than the Hawking-Page temperature, there exist new solutions that can be obtained by choosing 2nd or 3rd or 4th solutions (\ref{G02-mu})-(\ref{G14-mu}) in the $q=2$ trial function. However, all such solutions that we find have action higher than that of the standard saddle point. 
	\item The Maldacena-Qi hysteresis \cite{Maldacena18} of the free energy is reproduced by our approach, as demonstrated by the blue curve on Fig.\ref{fig:q=4-hysteresis}. In terms of the trial functions, the $q=4$ solutions on the horizontal branch of the hysteresis are obtained from $q=2$ solutions with large enough number of 2nd roots among Matsubara modes (in our case $14$). The solutions on the linearly decaying branch are obtained from the trial function with the 1st root for all Matsubara modes. 
	\item We have also found that at low temperatures the saddle points with $G^{(2)}$-solutions in the trial function do not exist. They appear around the Hawking-Page transition, and have higher free energy, which is moreover increases with temperature. On Fig.\ref{fig:q=4-hysteresis} the red curve is the free energy of a $q=4$ saddle point, obtained from the $q=2$ solution with $G_{\alpha\beta}(\pm \omega_0) = G^{(2)}_{\alpha\beta}(\pm \omega_0)$, which corresponds to the phase \textbf{0} in the $q=2$ model. 
\end{itemize}

Thus we conclude that the phase diagram of the traversable wormhole model with $q=4$ is drastically different from its $q=2$ counterpart, all due to the Hawking-Page phase transition, which does not happen in the $q=2$ case. This is a key distinction between the traversable wormhole model and our model with nonlocal symmetric coupling. More specifically, the model with the interaction (\ref{eta_sym(t)T}) promotes the subleading saddle points of the two decoupled SYK replicas into non-trivial phase structure, whereas the model dual to the traversable wormhole cannot do this due to the Hawking-Page transition. 

\section{Symmetry breaking and quasi-averaging}
\label{sec:Symmetry}

This section is devoted to the discussion of patterns of continuous symmetry breaking in two SYK replicas with and without interaction. 

\subsection{Spontaneous symmetry breaking in two decoupled replicas}

When the interaction is turned off $\eta =0$, the path integral (\ref{Z(eta)}) at finite temperature is invariant under the time translations in each replica $U(1) \times U(1)$. The replica-diagonal saddle points preserve this symmetry. However, the replica-nondiagonal saddle points in general spontaneously break this symmetry. In particular, the saddle points, for which $G_{\alpha\beta}$ depend only on difference of times, spontaneously break the full time translation symmetry as $U(1) \times U(1) \rightarrow U(1)$. This has important consequences, in particular in \cite{Saad18} it was shown that this symmetry breaking in the spectral form factor explains the ramp behavior. 

Because of the spontaneous symmetry breaking, one can generate other replica-nondiagonal solutions by acting with a broken generator. The broken symmetry acts on the components of solution for the field $G(\tau_1 - \tau_2)$ in the coordinate space as follows:
\bea
&& G_{LL}(\tau) \to G_{LL}(\tau)\,; \qquad G_{RR}(\tau) \to G_{RR}(\tau)\,;\\
&& G_{LR}(\tau) \to G_{LR}(\tau + \alpha)\,; \qquad G_{RL}(\tau) \to G_{RL}(\tau + \alpha)\,, \label{U(1)}
\eea 
and in the frequency space as
\bea
&& G_{LL}(\omega) \to G_{LL}(\omega)\,; \qquad G_{RR}(\omega) \to G_{RR}(\omega)\,;\\
&& G_{LR}(\omega) \to \e^{-i \omega \alpha} G_{LR}(\omega)\,; \qquad G_{RL}(\omega) \to \e^{+i \omega \alpha} G_{RL}(\omega)\,. \label{U(1)f}
\eea
One can check for generic $q$ that the on-shell action (\ref{Ion-shell2}) is invariant under these transformations, provided $\eta = 0$. Let us now make a few remarks: 
\begin{itemize}
	\item Since in the decoupled case the replica-nondiagonal solutions are subleading at large $N$, we use here the term "spontaneous symmetry breaking" in a generalized sense. It does not imply the existence of an ordered phase with spontaneously broken symmetry in the thermodynamic limit $N \to \infty$, but takes into account subleading saddle points as well.
	\item An important particular transformation of the form (\ref{U(1)}) is realized by setting $\alpha=\frac{\beta}{2}$. It can be used to map replica-symmetric solutions in the SYK with $M=2$ decoupled replicas to replica-antisymmetric solutions, which are related to thermofield double correlators \cite{AKTV}. This simple mapping works only for $M=2$ replicas, but in principle for general $M$ one can generate other solutions, starting from the replica-symmetric solutions or solutions with Parisi pattern of replica symmetry breaking. 
	\item For $q=2$ the Matsubara modes are decoupled. Therefore, the symmetry (\ref{U(1)f}) can be treated as gauge symmetry with $\alpha = \alpha(\omega_n)$. 
\end{itemize}

\subsection{Quasi-averaging}

The interaction between the replicas works as a source term in the bilocal replica field action. As we mentioned in the end of section \ref{sec:Setup}, it turns the spontaneous breaking of the time translation symmetry into the explicit one. In other words, the source $\eta$ lifts the degeneracy between the replica-nondiagonal solutions of the decoupled system. In the sections \ref{sec:q=2} and \ref{sec:q=4} we have shown that the source of the particular form (\ref{eta_sym(t)T}) generates a nontrivial phase structure from the replica-nondiagonal solutions which have the lowest free energy after the split of the degeneracy. To work towards the physical implications of these effects, it is important to define a quantity that is sensitive to the lifting of the degeneracy. 

In the SYK with two decoupled replicas, the exact two-point correlator is defined as 
\be
\mG_{\alpha\beta}(\tau_1, \tau_2) = \left.\frac{\delta}{\delta \eta_{\alpha\beta}(\tau_1, \tau_2)} \log \mZ(\eta) \right|_{\eta=0}
\ee
We are specifically interested in the case when $\alpha\neq \beta$. In terms of the path integral (\ref{Z(eta)}), we have 
\be
\mG_{\alpha\beta}(\tau_1, \tau_2) = \frac{1}{\mZ(0)}\int DG D\Sigma \e^{-N I[G, \Sigma; \eta]} G_{\alpha\beta}(\tau_1, \tau_2)|_{\eta = 0}\,.
\ee
We would like to study this quantity in thermodynamic limit $N \to \infty$. However we have the spontaneous symmetry breaking, which is made explicit by the source $\eta$. Therefore, one has to pay special attention to the order of taking the limits $N \to \infty$ and $\eta \to 0$ \cite{Bogolyubov1}. 

One can define the usual quantum average in thermodynamic limit
\be
\langle G_{LR}(\tau_1, \tau_2) \rangle = \lim_{N \to \infty} \left[\frac{1}{\mZ(0)}\int DG D\Sigma \e^{-N I[G, \Sigma; \eta]} G_{LR}(\tau_1, \tau_2)|_{\eta = 0} \right]\,.
\ee
To compute this quantity, one sets $\eta = 0$ first and then evaluates the path integral by the saddle point. The dominant saddle for $\eta = 0$ is the standard replica-diagonal solution, which preserves the time translation symmetry. Therefore, the result is 
\be 
\langle G_{LR} \rangle = 0\,.
\ee

The more suitable quantity for probing the replica-nondiagonal structure is the Bogolyubov \textit{quasi-average} \cite{Bogolyubov1,Arefeva19}. In our case we define it as follows: 
\be
\prec G_{LR}(\tau_1, \tau_2) \succ = \lim_{\eta \to 0} \left[\lim_{N \to \infty} \frac{1}{\mZ(0)}\int DG D\Sigma \e^{-N I[G, \Sigma; \eta]} G_{LR}(\tau_1, \tau_2) \right]\,. \label{quasi-average}
\ee
To compute this quantity, one has to evaluate the path integral by the saddle point at some non-zero $\eta$ and then take the limit $\eta \to 0$. This quantity is multivalued: its value depends on the initial value of $\eta$, since it determines the dominant saddle point. This allows this quantity to probe the replica-nondiagonal structure

Let us demonstrate how this works on the $q=2$ version of the model (\ref{Stotal}). First, we take a finite value of $\nu=\nu_0$ such that the system sits in the phase \textbf{0}, i.e. $G_1(\pm \omega_0) = G_1^{(4)}(\pm \omega_0)$ (see \ref{G14-nu}). In this case 
\be
\lim_{N \to \infty} \frac{1}{\mZ(0)}\int DG D\Sigma\ \e^{-N I[G, \Sigma; \eta]} G_{LR}(\omega_0)= G_1^{(4)}(\omega_0)\,.
\ee
Now we have to take the limit $\nu \to 0$. For this we use the expansion (\ref{G14-nu-ser}):
	\be
	G_1|_{\nu=0}(\omega) =i \sgn(\omega) \frac{\sqrt{4J^2 + \omega^2}}{2J^2}\,.
	\ee
As a result, this gives the non-zero quasi-average 
\be
\prec G_{LR}(\omega_0) \succ =i \frac{\sqrt{4J^2 + \omega_0^2}}{2J^2}\,; \qquad \prec G_{LR}(\omega_n) \succ = 0 \quad \forall n \neq 0\,.
\ee
The value of the quasi-average $\prec G_{LR}(\tau_1, \tau_2) \succ$ matches the value of the replica-offdiagonal component of $G_{\alpha\beta}$ on the replica-nondiagonal saddle point of the SYK with decoupled replicas, which can be found analytically for $q=2$ and numerically for $q=4$ \cite{AKTV}. One can say that the property of the model (\ref{Stotal}) is that the quasi-averages remember about its phase structure after $\nu$ is turned off. Note that this is not the case in the model with local coupling (\ref{zeta_antisym(t)}): in both $q=2$ and $q=4$ versions of the model the dominant phases have $G_1|_{\mu = 0} = 0$. 

As a final remark, let us note that originally quasi-averages were proposed by Bogolyubov as a probe of ordered phases with spontaneously broken symmetry \cite{Bogolyubov1}. Our considerations in \cite{AKTV} and in the present paper show that the two-replica SYK is different in this regard: there are no phases with spontaneously broken symmetry, but there is still spontaneous symmetry breaking in the sense of non-zero quasi-averages. 

\section{Discussion}
\label{sec:Discussion}

We have studied the model of two SYK replicas with nonlocal interaction (\ref{Sint}). We have shown that in both $q=2$ and $q=4$ cases this model exhibits a nontrivial phase structure with infinite number of phases. The coupling of the form (\ref{eta_sym(t)T}) is unique, because every nontrivial phase corresponds to a replica-nondiagonal saddle point in the SYK with two decoupled replicas. This mapping is made explicit by consideration of quasi-averages, which correspond to the breaking of time translation symmetry. 

Let us now summarize the main results of the paper: 
\begin{itemize}
	\item[1.] We have solved the saddle point equations of two nonlocally coupled SYK$_2$ chains. We have derived the phase diagram (\ref{Tcr-sym}), see Fig.\ref{fig:q=2-PD-sym}. We have shown that the nontrivial phases correspond to replica-nondiagonal saddle points of the two-replica SYK$_2$ when the coupling is turned off. 
	\item[2.] We have found numerical solutions of saddle point equations of two nonlocally coupled SYK$_4$ chains, see Figs.\ref{fig:sol-nu=01},\ref{fig:sol-nu=5}. We have shown that this model demonstrates nontrivial phase structure similar to the $q=2$ counterpart. We have studied the behavior of the annealed free energy, see Fig.\ref{fig:q=4-phase-transitions}. 
	\item[3.] We have solved the saddle point equations of two locally coupled SYK$_2$ chains. We have also obtained the phase diagram for that model, see Fig.\ref{fig:q=2-PD-antisym}. In this case, all phases, including the nontrivial ones, correspond to replica-diagonal saddle points in two-replica SYK$_2$, when the coupling is turned off. 
	\item[4.] We have performed a numerical study of saddle point structure of the model of two locally coupled SYK$_4$ chains, which is dual to the traversable wormhole. We have found that the phase structure is drastically different from the $q=2$ case, due to effects related to the Hawking-Page phase transition. We have reproduced the Maldacena-Qi hysteresis of the free energy by a different approach, and we have shown that other saddle points have higher free energy than the Hawking-Page saddles. 
	\item[5.] We have discussed the connection between the model of two nonlocally coupled SYK chains with the spontaneous symmetry breaking of time translation symmetry by replica-nondiagonal solutions in the two-replica SYK. From analytic arguments in $q=2$ model and numerical solutions in $q=4$ model, we see that thermodynamic limit $N \to \infty$ does not commute with the limit of zero interaction between replicas $\nu \to 0$. Because of this, the replica-offdiagonal correlator is non-zero in the sense of quasi-averages. This establishes the spontaneous symmetry breaking in the two-replica SYK in the sense of non-zero quasi-averages. 
\end{itemize}
%

The connection between the phases of the model (\ref{Stotal}) and the quasi-averages in the decoupled SYK replicas motivate us to propose this model as a tool for probing the nonperturbative effects in two decoupled SYK replicas. One can do this by performing the following operations: 
\begin{itemize}
	\item Prepare an equilibrium configuration of the model (\ref{Stotal}) by tuning the coupling $\nu$ and temperature $T$ such that the system is in the phase ${\bf n}_0$. 
	\item Turn off the coupling between replicas $\nu$ instantly. 	
\end{itemize}
After such quantum quench we end up with the two-replica SYK in a non-equilibrium configuration with non-zero value of the replica-nondiagonal correlator. This non-equilibrium configuration is described by a subleading saddle point. It can be expected that this configuration will live for a time of order 
\be
\Delta t \sim \frac{\beta}{N \delta}\,,
\ee
where 
\be
\delta = I (\text{replica-nondiagonal}) -  I (\text{standard})
\ee
is the difference between the on-shell action of the replica-nondiagonal and the standard replica-diagonal saddles. Until this configuration decays, we can expect that the physics of the system at large $N$ will be determined by the replica-nondiagonal saddle point, which is usually a nonperturbatively small effect in the equilibrium state. Thus, the model of two nonlocally coupled SYK chains (\ref{Stotal}) provides a protocol of a quantum quench, which can turn a small nonperturbative effect in regular SYK into a non-equilibrium configuration which dominates the physics for a time period of order $1/N$. The computation of real-time nonequilibrium correlators in such quench scenario is an interesting problem for future work. 

The caveat to this problem, which might make the physical realization of such quench protocol difficult, is the fact that the initial state is prepared as a phase with negative heat capacity, as we discuss in section \ref{sec:q=4-phases}. The negative heat capacity might hint that the model (\ref{Stotal}) as it is could not be realized physically, but one can probably cure this by turning the source $\nu$ into a dynamic variable with some slow enough dynamics. 

The other pressing open question is to find out what is the precise role of the replica-nondiagonal saddle points and of the coupling (\ref{Sint}) in the dual gravity description. We expect that the replica-nondiagonal saddle points in SYK with decoupled replicas, obtained in \cite{AKTV} describe the contributions of nontrivial topologies to the path integral of UV completion of the JT gravity \cite{gravity}, which were discussed in \cite{Saad19}. The work \cite{Maldacena17,Maldacena18} has shown that the model with local interaction defined by (\ref{zeta_antisym(t)}) is suitable as a holographic dual to traversable wormhole. From our study of the model with nonlocal interaction (\ref{eta_sym(t)T}), we can say that the holographic dual is different from a wormhole. It would be interesting to check whether our model is perhaps more related to some D-brane configurations in the conjectured theory of JT string instead. Also, the negative heat capacity can motivate to consider the holographic dual model, where some form of the Hawking evaporation happens.

\section*{Acknowledgements}

The authors are grateful to Maria Tikhanovskaya and Valentin Zagrebnov for useful discussions. M. K. is supported by the Foundation for the Advancement of Theoretical Physics and Mathematics ``BASIS'' (project 17-15-566-1).

\end{document}